\documentclass[12pt,reqno]{amsart}
\usepackage{amssymb,latexsym,amsmath}
\usepackage{amscd,amsfonts}
\usepackage{epsfig}

\newtheorem{theorem}{Theorem}
\newtheorem{lemma}{Lemma}
\newtheorem{proposition}{Proposition}
\newtheorem{corollary}{Corollary}
\newtheorem{definition}{Definition}
\theoremstyle{definition}
\newtheorem{remark}{Remark}
\newtheorem{example}{Example}
\numberwithin{equation}{section}
\def\R{\mathbb{R}}

\begin{document}
\title[Curvature of the Weinhold Metric]{Curvature of the Weinhold Metric for thermodynamical
 Systems with 2 degrees of freedom}
 \date{April 4, 2005}
\author{Manuel Santoro}\address{Department of Mathematics and Statistics,
 Portland State University, Portland, OR, U.S.} \email{emanus@pdx.edu}
\author{Serge Preston}\address{Department of Mathematics and Statistics,
Portland State University, Portland, OR,
U.S.}\email{serge@mth.pdx.edu}

\begin{abstract}
In this work \footnote{} the curvature of Weinhold
(thermodynamical) metric is studied in the case of systems with
two thermodynamical degrees of freedom.  Conditions for the Gauss
curvature $R$ to be zero, positive or negative are worked out.
Signature change of the Weinhold metric and the corresponding
singular behavior of the curvature at the phase boundaries are
studied. Cases of systems with the constant $C_{v}$, including
Ideal and Van der Waals gases, and that of Berthelot gas are
discussed in detail.
\end{abstract}\bigskip
\maketitle
\section{Introduction}
Usage of geometrical methods in homogeneous Thermodynamics started
in the works by J.Gibbs (\cite{G}) and C.Caratheodory (\cite{Ca})
was further developed in the works of R.Hermann, R.Mrugala, in the
dissertation of H.Heemeyer (\cite{He}) and in other works.
Thermodynamical metrics also have their source in the works of J.
Gibbs (\cite{G}). Explicitly a thermodynamical metric (TD-metric)
was introduced by F.Weinhold (\cite{W}) and, later, from a
different point of view, by G.Ruppeiner (\cite{R}). Deeper studies
by P.Salamon and his collaborators, by P.Mrugala and H.Janyszek
(see \cite{BSI,J1,J2,JM1,JM2,M2,M3,M4,NS,SNI}) clarified principal
properties of thermodynamical metrics, relations between different
TD-metrics and their place in relation to the contact structure of
equilibrium thermodynamical phase space (\cite{H,M}). G. Ruppeiner
(see review \cite{R2} and the bibliography cited there) has
developed a covariant thermodynamical fluctuation theory based on
the Riemannian metric $\eta_{S}$ defined by the second momenta of
entropy with respect to the fluctuations and related the curvature
of this metric to the correlational volume near the critical
point.  He applied this scheme to a variety of models: ideal gas,
ideal paramagnetic, Ising model, Takahashi gas, Van der Waals gas.
P.Mrugala, H. Janyszek, P.Salamon studied the role of
thermodynamical metrics in the statistical thermodynamics, one of
examples being the model magnetic system (\cite{JM2}).\par
 These and some other developments, including an interest to
thermodynamical metrics in the study of black holes led us to the
question of a more systematic study of the curvature of
 thermodynamical metrics in the case of systems of smallest
dimension where such curvature is present - the case where the
phase space of TD-system is one with variables
$(U;(T,S),(p,V),(\mu,N))$. Degeneracy of a TD metric $\eta$ due to
the homogeneity of a constitutive law (see \cite{SNI} or see
below) is removed by considering 1 mole of the media (or by the
reduction procedure, \cite{P}). As a result, a TD-metric is
defined on the 2D constitutive surface (containing all equilibria
states of the system). All the intrinsic curvature properties of
this metric are defined by the scalar (Gauss) curvature $R$.
\par

 In Section 2 we introduce the thermodynamical metrics
$\eta_{E}$ defined by a thermodynamical potential $E$ on the
corresponding "constitutive surfaces" in the phase space of a
thermodynamical system endowed with the contact structure
(\cite{C,H}).\par In Section 3 we introduce the elementary
thermodynamical system (of two TD degrees of freedom) and
calculate its Weinhold and Ruppeiner metrics.  In addition we
present examples of TD-metrics of a chemical system and the
systems with magnetic and electrical properties.\par

 In section 4
we calculate Levi-Civita connection and the curvature tensor of
the metric $\eta_{E}$ in the general n-dim case(see also
\cite{R3}). In section 5 we introduce and study the Hessian
surface $H_{E}$ of the TD potential $E$ in the space of
symmetrical $2\times 2 $ matrices and express the curvature
properties of the metric $\eta_{E}$ in terms of geometrical
properties of this surface.  In particular, we determine the
condition for the positivity, negativity and nullity of the scalar
curvature $R(\eta_{E})$ in terms of this surface.\par

 In section 6 we continue the study in
of curvature started in section 4 by calculating the determinant
$det(\eta_{E})$ (characterizing the signature of the metric) and
the scalar curvature of the Weinhold metric $\eta_{U}$ (defined by
the internal energy $U$) of an elementary thermodynamical system
in terms of $T,V$, heat capacity $C_{v}$, volume coefficient of
expansion $\alpha $, isothermal compressibility $k$ and their
derivatives. \par In Sec.7 we consider the case of {\bf constant
$C_{v}$} where calculations simplify greatly. We define the form
of the fundamental constitutive relations $U=U(S,V)$ in such a
case and determine the systems of this type for which $R(\eta)=0$
and the regions where scalar curvature is positive or
negative.\par

 In Sec.8, example of the ideal gas is considered.
In Sec.9, we study the Weinhold metric for the Van der Waals gas,
curve of the signature change (degeneracy curve of the $det(\eta
)$), separating the domain of (stable) equilibria from the
unstable region, scalar curvature behavior near this curve,
calculation of critical values of physical variables in terms of
the degeneracy curve.   In Sec.11 the case of Berthelot gas is
considered.\par

  In the last Section 12 the geodesic equations of
the Weinhold metric $\eta_{U}$ for the general elementary
thermodynamical system are obtained.

 \mathstrut

\section{Weinhold Metric of a thermodynamical system}
In this section we introduce, following  \cite{BSI,H,M,M4}, some
geometrical structure of on the (energy-phase) state space of
homogenous thermodynamics and, in particular, the thermodynamical
metrics (\cite{W}) on the constitutive surfaces of this system -
the principal object of our study.
\par We introduce the {\bf thermodynamical phase space} $P$ of a
thermodynamical system with $k$ "thermodynamical degrees of
freedom" as a $(2k+1)$-dimensional manifold endowed with a contact
structure - 2D-distribution $D$ on $P$ which is totally
non-integrable, \cite{A}. This structure is presented (locally,
and in many cases globally) by a contact (Pfaff) form $\omega $
such that  $\omega \wedge (d\omega )^{k}\ne 0 $ at each point of
$P$.\par

 The fundamental geometrical structure  of a thermodynamical system
   is a choice of
a canonical chart (Darboux chart) $\mathcal C$ of variables
$z=(E,x^{i},y_{i},i=1,\ldots k)$ for the 1-form $\omega$,
realizing it in the form
\begin{equation}
\omega = dE-\sum_{i=1}^{k}y_{i}dx^{i}.
\end{equation}
Couples of variables $(x^{i},y_{i})$ have the meaning of {\bf
extensive} ($x^{i}$) and {\bf intensive} ($y_{i}$) variables,
corresponding to the different processes that may undergo in the
system and $E$ is the chosen thermodynamical potential (internal
energy, entropy, Helmholtz free energy or entalpy, see \cite{C}).
Examples of couples $(x^{i},y_{i})$ are: 1) temperature and
entropy $(T,S)$; 2) pressure and volume $(-p,V)$; 3) mole number
of $i$-th component and corresponding electrochemical potential
$(N_{i},\mu _{i})$; etc.\par

Horizontal distribution $D$ of contact structure - subbundle of
the tangent bundle $T(P)$
\begin{equation}
D_{z}=Ker(\omega (z)),\ z\in P
\end{equation}
is endowed with the symplectic form $d\omega $ having, in a
canonical chart $\mathcal C$, the standard form
\begin{equation}
d\omega= \sum_{i=1}^{k}dy^{i}\wedge dx^{i} \end{equation}
\par

Admissible processes described by the system are {\bf horizontal}
curves $\gamma : t\rightarrow z(t)\in P$ of structure $\omega $
(i.e. such that $\gamma'(t)\in D_{\gamma(t)}$ for all $t$). In the
case where $E$ is interpreted as the internal energy of the
system, $\int_{\gamma}dE$ is the work applied (or produced) during
the process $\gamma (t)$, see (\cite{C,KP}), and condition of
admissibility ensures the fulfillment of the energy conservation
law during the process.\par
 {\bf Constitutive surfaces} are
defined as maximal integral surfaces of structure $\omega $
(Legendre submanifolds), (\cite{A,M}). Standard way to present
such a surface $\Sigma_{E}$ is to define it as the first jet of a
function $E=E(x^{i},i\in I;y_{j},j\in  [1,\ldots ,k]\backslash I
)$, defined in the open subset $D\subset R^{k}$. Here we choose a
subset $I\subset [1,\ldots ,k]$ of indices and take $E$ to be a
function of $x^{i}$ with $i\in I$ and of $y_{j}$ with $j$ in the
complemental set of indices $j\in [1,\ldots ,k]\backslash I$.
Relation $E=E(x^{i},y_{j})$ defines the fundamental constitutive
relation of TD system, (see \cite{C}), from which all others
equations of state follows (\cite{C}). In the usually considered
case where $I=[1,\ldots , k]$ corresponding constitutive surface
is given by
\begin{equation}
\Sigma_{E}=\{(E,x^{i},y_{j},i=1,\ldots k)\in P\vert E=E(x^{i}),
y_{i}=\frac{\partial E}{\partial x^{i}},i\in [1,\ldots , k],\
(x^{i}) \in D\subset R^{k}\}.
\end{equation}
\par
 From now on we will denote arguments of a function $E$ defining
a fundamental constitutive relation and the corresponding
equilibria surface by $x^{i},\ i\in [1,\ldots k]$.
\par
\begin{remark} Constitutive surface $\Sigma_{E}$ defined by a
function $E(x^{i})$ contains the subsets ${\mathcal Eq}_{C}
\subset \Sigma_{E}$ of {\bf equilibria states} of the homogeneous
thermodynamical system $E=E(x^{i})$ subject to the (possible)
constraints imposed on the system in the situation $C$ (see
\cite{C} or other standard text in the thermodynamics for a
variety of such situations). Subset ${\mathcal Eq}_{C}\subset
\Sigma_{E}$ is determined, due to the second law of
thermodynamics, by the condition of extremum of the
thermodynamical potential $E$ subject to the constraints $C$, see
\cite{C,KP}.  Submanifold $\Sigma_{E}$ contains also unstable
states as well as {\it locally stable} states (\cite{C}).
Partition of the constitutive surface $\Sigma_{E}$ corresponding
to a given family of constraints was studied in many cases (see
\cite{ML,C}). Geometry of the constitutive surface $\Sigma_{E}$,
of its equilibria region(s) ${\mathcal Eq}_{C}$ in relation to the
admissible thermodynamical processes is one of the primary object
of study of "geometrical thermodynamics", see \cite{ML,M,M3,M4}).
\end{remark}

Thermodynamical metric (TD-metric) defined by the constitutive
relation $E=E(x^{i})$ on the constitutive surface $\Sigma_{E}$ of
the contact structure $\omega$ has the form
\begin{equation}
\eta _{E}=\sum_{ij}\frac{\partial^{2} E}{\partial x^{i}\partial
x^{j}}dx^{i}\otimes dx^{j}.
\end{equation}
For the case where $E$ is the internal energy $U$, metric
$\eta_{U}$ is called the Weinhold metric. For the case where $E$
is the entropy $S$, instead, metric $\eta_{S}$ is called the
Ruppeiner metric.
\par
 Notice that
this form of the metric is not invariant under the diffeomorphisms
of domain $D$ (except linear ones, see \cite{BSI} for more). Thus,
though is not a drawback, TD-metric is defined by a physically
meaningful choice of canonical chart $\mathcal C$ and
transformation of this metric under the transition from one
physically meaningful canonical chart to another one cannot be
simple.  As an example, we mention that the Legendre
transformation of the space $P$ leading to the replacement of
internal energy $U$ by the entropy $S$ applied to the TD-metric
$\eta_{U}$ gives, as is shown in \cite{M3,SNI}, the TD-metric,
conformal to the metric $\eta_{S}$ on the equilibria surface
$X_{U}$,
\[ \eta_{U}=-\frac{1}{T}\eta_{S}.
\]
\par
Metric $\eta_{E}$ with $E=U$ being the {\bf internal energy} of a
thermodynamical system, was introduced by F.Weinhold in series of
papers \cite{W}. It was further studied by P.Salaman, S.Berry and
their collaborators (see \cite{BSI,NS} and the bibliography in
\cite{PV}). \par

\begin{remark} A TD-metric $\eta_{E}$  is induced
on the surface $\Sigma_{E}$ of the form (2.1) by the following
symmetrical tensor
\begin{equation}
{\tilde \eta}=\frac{1}{2}\sum_{i=1}^{k}(dy_{i}\otimes
dx^{i}+dx^{i}\otimes dy^{i}).
\end{equation}

This tensor is the only (up to a conformal factor) symmetrical
tensor in $P$, annihilating the Reeb vector field $Y$ of structure
$\omega $ (here $Y=\frac{\partial }{\partial E}$), invariant under
the substitution of indices $i$, and obtained as the sum of
symmetrical tensors in the 2D-elementary subspaces $D_{i}^{*}$ of
$D_{x}^{*} $ spanned by couples of covectors $(dx^{i},dy_{i})$ of
thermodynamically conjugated variables.
\end{remark}

A very interesting way to construct all the TD-metrics $\eta_{E}$
in the framework of the contact geometry was suggested by
R.Mrugala (\cite{M2}). He has defined (starting from some
arguments of statistical mechanics) the contact metric $G$ on the
phase space $P$ inducing TD metric (2.5) on each equilibria
surfaces.  Geometrical properties of this metric are studied in
the forthcoming work of the second author with J.Vargo \cite{PV}
.\par

Function $E$ defining constitutive relation and the constitutive
surface $\Sigma_{E}$ is assumed to be homogeneous of the first
order: $E(\lambda x^{i})=\lambda E(x^{1})$,\cite{C}.  As a result,
Weinhold metric is degenerate - its kernel is generated by the
radial vector field in the space $X$
\[
Ker (\eta )=R\cdot (x^{i}\frac{\partial }{\partial x^{i}})
\]
This can be remedied either restricting all considerations to the
subspace of $X$ (and the section of cone $\Sigma_{E}$) with fixed
value of one of variables $x^{i}$ (mole number or volume are
typical examples, see \cite{NS}) or using the geometrical
reduction of the surface $\sigma_{E}$ by the action of this
dilatation group (\cite{P}). \par

After this reduction of the TD-system its Weinhold metric is in
general non-definite - the constitutive surface $\Sigma_{E}$ is
the union of domains of where this metric has different signature
separated by the submanifolds (generically of codimension one) of
states where this metric is degenerate. Submanifold of
thermodynamical equilibria in the processes with a given set of
constrains $C$ - ${\mathcal Eq}_{C}$ of the constitutive surface
$\Sigma_{E}$ lays the region of the {\bf definite TD-metric} due
to the second law of thermodynamics which requires the entropy to
be (locally) maximal and energy to be (locally) minimal in the
stable equilibria state to which system tends.
\par

Gauss curvature $R$ of thermodynamical metrics (in energy or
entropy form) was calculated for several TD systems with two
degrees of freedom (i.e. 5-dim phase space $P$) - ideal gas
($R=0$), Takahashi gas, multi-component ideal gas, paramagnetic
ideal gas, Van-der-Waals gas, ideal quantum gases(see review
\cite{R}). G. Ruppeiner (see review \cite{R2} and the bibliography
cited there) related the curvature $R(\eta_{S})$ of the metric
defined by entropy to the correlational volume near the critical
point.  The fact that for the ideal gas the curvature of Weinhold
metric $\eta_{U}$ and that of Ruppeiner metric $\eta_{S}$ is zero
and the calculations of curvature for systems mentioned above
allows to suggest some relation between this curvature (or at
least its sign or nullity) and the interactions undergoing in the
system on the microscopic level.  Clarification of these questions
requires deeper study of intrinsic and especially extrinsic
geometry of constitutive submanifolds $\Sigma_{E}$ of the
thermodynamical phase space $P$.

\section{Elementary TD system and other examples}
Here we present several examples of TD-metrics including the basic
{\bf elementary thermodynamical system} with 5-dim phase space $P$
in variables $(U,S,T,p,V).$\par

\subsection{Elementary TD-system}
As the first example of a thermodynamical system and its Weinhold
and Ruppeiner metrics we consider the system with  5-dimensional
phase-energy space  of variables $(U,(S,T),(V,-p))$ which we will
call from now on an {\bf elementary thermodynamical system}. This
system will be the principal object of our study here, so we will
consider it in more details. Contact form $\omega $ of such a
system can be chosen in a form
\begin{equation}
\omega =dU-TdS+pdV.
\end{equation}

 Consider a 2D integral submanifold $\Sigma_{U}$ of this system
defined by a constitutive relation
\begin{equation}
U=U(S,V),
\end{equation}
with $S$ and $V$ being our extensive variables, $T$ and $p$ the
relative conjugate intensities and $U$ the internal energy
function of the extensive variables.
\par
Then
\begin{equation}
dU=(\frac{\partial U}{\partial S})dS+(\frac{\partial U}{\partial
V})dV =TdS-pdV.
\end{equation}
Notice that homogeneity condition for any constitutive relation in
U-system (\cite{C}) leads to the relation
\begin{equation}
U=ST-pV,
\end{equation}
that should be fulfilled for all integral surfaces of contact
structure $\omega $ given by a constitutive equation.
Correspondingly, contact condition for thermodynamical processes
has two forms, i.e.
\[
\omega =dU-TdS+pdV= SdT-Vdp=0
\]
where the second one (Gibbs-Duhem equation, see \cite{C} ) leads
to the second form of constitutive relation provided the
constitutive relation in the first (fundamental) form is
chosen.\par

 Calculating
the Weinhold metric of U-system we will use some notations
standard in the literature (see \cite{C}). Namely, we denote
\begin{enumerate}
\item $C_{v}$ is the heat capacity at constant volume:
\begin{equation}
C_{v}=T(\frac{\partial S}{\partial T})_{V},
\end{equation}
\item $C_{p}$ is the heat capacity at constant pressure:
\begin{equation}
C_{p}=T(\frac{\partial S}{\partial T})_{p},
\end{equation}
\item $\alpha$ is the volume coefficient of expansion:
\begin{equation}
\alpha=\frac{1}{V}(\frac{\partial V}{\partial T})_{p}
\end{equation}
\item $k$ is the isothermal compressibility:
\begin{equation}
k=-\frac{1}{V}(\frac{\partial V}{\partial p})_{T}
\end{equation}
\end{enumerate}
These quantities are related by:
\begin{equation}
C_{p}-C_{v}=VT\frac{\alpha^{2}}{k}
\end{equation}

Calculating Weinhold metric for U-system on the two-dimensional
integral surface $\Sigma_{U}(S,V)$ and using introduced notations
we get

\begin{equation}
\eta_{U} =(\eta_{ij} )=\frac{1}{C_{v}}
\begin{pmatrix}
 T & -\frac{T\alpha}{k}\\
 -\frac{T\alpha}{k} & \frac{C_{p}}{vk}
\end{pmatrix}
\end{equation}
Characteristic equation of tensor $\eta_{ij}$ (with respect to the
canonical Euclidian metric $h$ on the $(S,V)$-plane) is

\begin{equation}
\lambda^{2}-(\frac{T}{C_{v}}+\frac{C_{p}}{C_{v}Vk})\lambda+\frac{T}{C_{v}Vk}=0,
\end{equation}
\par
and
\[
\lambda_{\pm}=\frac{1}{2}[(\frac{T}{C_{v}}+\frac{C_{p}}{C_{v}Vk})\pm
\sqrt{\triangle}]
\]
For the discriminant $\triangle $ we have the following inequality
\par
\begin{equation}
\triangle=(\eta_{11}-\eta_{22})^{2}+(2\eta_{12})^{2}=
(\frac{T}{C_{v}}-(\frac{C_{p}}{C_{v}Vk}))^{2}+(2\frac{T\alpha}{C_{v}k})^{2}>0
\end{equation}
\par
which, in particular, implies that
\par
\begin{equation}
\det_{h}\eta_{ij}<\frac{1}{4}(trace_{h}(\eta_{ij}))^{2}
\end{equation}
\par
\begin{remark}
Notice that since $C_{p}=C_{v}+\frac{VT\alpha^{2}}{k}$,
discriminant $\triangle$ is always positive. Therefore eigenvalues
$\lambda_{\pm}$ are both real and distinct.
\end{remark}
\par
Moreover, since $\det\eta_{ij}=\lambda_{+}\lambda_{-}$, and since
also $\det\eta_{ij}=-\frac{T}{C_{v}}(\frac{\partial p}{\partial
V})_{T}$, we get the following
\par
\begin{lemma}
Let $T>0$. Then
\par
$1)$ If $C_{v}>0$ and $(\frac{\partial p}{\partial V})_{T}<0$ then
$\det\eta_{ij}>0$ and $\lambda_{\pm }>0$.
\par
$2)$ If $C_{v}<0$ and $(\frac{\partial p}{\partial V})_{T}>0$ then
$\det\eta_{ij}>0$ and $\lambda_{\pm }<0$.
\par
$3)$ If $C_{v}$ and $(\frac{\partial p}{\partial V})_{T}$ have
different sign then $ \det\eta_{ij}<0$ and $\lambda_{+}>0$,
$\lambda_{-}<0$.
\end{lemma}
\par
Fulfillment of 1) and 2) guarantees the stability for the system
even if the case 2) might look non physical at all. However, there
are some cases when it is natural to assume that the heat capacity
is negative (in a cluster of sodium atoms or in the black holes,
\cite{PM}).  There are also cases in which the isothermal (and
adiabatic) compressibility seems to be negative (in amino acid,
\cite{MSS}). These are examples of cases away from stability.
\begin{example}
In case of an Ideal Gas, $\alpha=\frac{1}{T}$ and $k=\frac{1}{p}$,
and the metric $\eta_{U}$ is given by

\begin{equation}
\eta_{U ij} =\frac{1}{C_{v}}
\begin{pmatrix}
 T & -p\\
 -p & \frac{C_{p}p}{v}
\end{pmatrix}
\end{equation}
\end{example}

Applying Legendre's transformation from variables
$(U,(S,T),(V,-p))$ to the variables $(S,(U,T^{-1}),(V,pT^{-1}))$
to the contact form $\omega $ and rewriting the constitutive
relation correspondingly, we get

\begin{equation}
\omega = dU-TdS+pdV=-T\left[ dS-\frac{1}{T}dU-\frac{p}{T}dV\right]
=0\Rightarrow \omega =dS-\frac{1}{T}dU-\frac{p}{T}dV,
\end{equation}
\par
and
\par
\begin{equation}
S=S(U,V)
\end{equation}
\par
Correspondingly, TD-metric on the two-dimensional integral surface
$\Sigma_{S}(U,V)$ (Ruppeiner metric) takes the form, see also
\cite{M2},
\par
\begin{equation}
\eta_{S ij} =
\begin{pmatrix}
-\frac{1}{C_{v}T^{2}} & \frac{1}{T^{2}}(\frac{T\alpha-kp}{kC_{v}}) \\
\frac{1}{T^{2}}(\frac{T\alpha-kp}{kC_{v}}) &
-\frac{1}{C_{v}}(\frac{T\alpha-kp}{Tk})^{2}-\frac{1}{VkT}
\end{pmatrix}
\end{equation}
\par
In case of an Ideal Gas, metric (3.17)  takes the form
\par
\begin{equation}
\eta_{S ij}=
\begin{pmatrix}
-\frac{1}{C_{v}T^{2}} & 0\\
0 & -\frac{p}{VT}
\end{pmatrix}
\end{equation}
due to the Gibbs-Duhem form of constitutive relation
$(T\alpha-kp)=0$.
\par
\subsection{A chemical system}
Here we consider a (2m+5) homogeneous thermodynamical system in
energy-phase space defined by
\par
\begin{equation}
(S,(U,\frac{1}{T}),(V,\frac{p}{T}),(N_{1},\frac{\mu_{1}}{T}),...(N_{m},\frac{\mu_{m}}{T}))
\end{equation}
\par
with $N_{i}$ being the number of molecules of type $i$ and
$\mu_{i}$ being their conjugate intensities - chemical potential
per molecule i-th.
\par
Consider  (m+2)-dim integral surface defined by an entropy
constitutive relation
\par
\begin{equation}
S=S(U,V,N_{1},.....N_{m})
\end{equation}
\par
We get
\par
\[
dS=\left( \frac{\partial S}{\partial U}\right) dE+\left( \frac{%
\partial S}{\partial V}\right) dV+\sum_{i}\left( \frac{\partial S}{\partial
N_{i}}\right) dN_{i}
\]
\par
\begin{equation}
=\frac{1}{T}dU+\frac{p}{T}dV+\sum_{i}\frac{\mu _{i}}{T}dN_{i}.
\end{equation}
\par

\par
Then we calculate
\par
\begin{equation}
\eta_{S ij} =
\begin{pmatrix}
-\frac{1}{C_{v}T^{2}} & \frac{1}{T^{2}}(\frac{T\alpha-kp}{kC_{v}}) & \frac{%
\partial }{\partial E}(\frac{\mu _{j}}{T}) \\
 \frac{1}{T^{2}}(\frac{T\alpha-kp}{kC_{v}}) & -\frac{1}{C_{v}}(\frac{T\alpha-kp}{Tk})^{2}-\frac{1}{VkT} &
\frac{\partial }{\partial V}(\frac{\mu j}{T}) \\
\frac{\partial }{\partial N_{i}}(\frac{1}{T}) & \frac{\partial
}{\partial N_{i}}(\frac{p}{T}) & \frac{\partial }{\partial
N_{i}}(\frac{\mu _{j}}{T}).
\end{pmatrix}
\end{equation}

\vskip 1cm

 \subsection{Thermodynamical system in homogeneous Magnetic and Electric
 Fields.}
\par
Constitutive relation should now include the work of magnetic and
electric field on the (homogeneous) system. Let's consider the
work of magnetic field first. Fundamental result of the
thermodynamics of magnetic system defines the differential of the
magnetic work to be (\cite{C}):
\par
\begin{equation}
dW_{Magn}=d(\frac{\mu_{0}}{8\pi}\int{H^{2}}dV)+\mu_{0}\int(H\cdot
dM)dV,
\end{equation}
\par
where H is the external magnetic field, M is the magnetization
(magnetic momentum) vector. First term represents the change of
energy of the magnetic field, second, the work of the magnetic
field on the magnetic momentum of the system.  Since we consider
here only homogeneous systems, we take H to be constant. Let also
assume that the magnetization $M$ is homogeneous.
\par
Then
\par
\begin{equation}
dW_{Magn}=d(\frac{\mu_{0}}{8\pi}\int{H^{2}dV})+\mu_{0}H\cdot dI
\end{equation}
where $I^{tot}=\int MdV=MV$ is the total magnetic dipole moment of
the system.
\par
Considering just the contribution to the magnetic work due to the
magnetic dipole moment $I$, let's define the (internal) energy to
be
\par
\begin{equation}
U=U(S,V,I^{tot}_{H},N_{1},N_{2},....,N_{m}),
\end{equation}
where $I^{tot}_{H}$ is the component of the total magnetic moment
parallel to the external field. Then
\par
\[
dU=(\frac{\partial{U}}{\partial{S}})dS+(\frac{\partial{U}}{\partial{V}})dV+\sum_{i}(\frac{\partial{U}}{\partial{N_{i}}})dN_{i}
+(\frac{\partial{U}}{\partial{I^{tot}_{H}}})dI^{tot}_{H}
\]
\par
\begin{equation}
=TdS-pdV+\sum_{i}{\mu_{i}dN_{i}}+\mu_{0}HdI^{tot}_{H}
\end{equation}
where $\mu_{0}$ is the permittivity of free space.
\par
Then the Weinhold metric for the system with constitutive relation
(3.26) in the space of variables
$(S,V,I^{tot}_{H},N_{1},N_{2},....,N_{m})$ is
\par
\begin{equation}
\eta_{U ij}=
\begin{pmatrix}
  \frac{T}{C_{v}} & -\frac{T\alpha}{C_{v}k} & \frac{\partial{T}}{\partial{I^{tot}_{H}}} &
  \frac{\partial{T}}{\partial{N_{i}}} \\
  -\frac{T\alpha}{C_{v}k} & \frac{C_{P}}{vC_{v}k} & -\frac{\partial{p}}{\partial{I^{tot}_{H}}}
  & -\frac{\partial{p}}{\partial{N_{j}}} \\
  \mu_{0}\frac{\partial{H}}{\partial{S}} & \mu_{0}\frac{\partial{H}}{\partial{V}}
  & \mu_{0}\frac{\partial{H}}{\partial{I^{tot}_{H}}} & \mu_{0}\frac{\partial{H}}{\partial{N_{j}}} \\
  \frac{\partial{\mu_{i}}}{\partial{S}} & \frac{\partial{\mu_{i}}}{\partial{V}}
  & \frac{\partial{\mu_{i}}}{\partial{I^{tot}_{H}}}
  & \frac{\partial{\mu_{i}}}{\partial{N_{J}}}
\end{pmatrix}
\end{equation}
\par
Similarly, for the work of a homogeneous external electrical field
$E_{l}$ in homogeneous case we have (\cite{C})
\par
\begin{equation}
dW_{electr.}=d(\frac{\epsilon_{0}}{8\pi}\int{E^{2}_{l}}dV)+E_{l}dP^{tot}
\end{equation}
where $P^{tot}=PV $ is the total electric dipole moment.
\par
Similarly to the previous case, if we consider just the
contribution to the electric work due to the electric dipole
moment, we can define the internal energy as follows
\par
\begin{equation}
U=U(S,V,P^{tot}_{E},N_{1},.....N_{m})
\end{equation}
Then
\par
\begin{equation}
dU=TdS-pdV+E_{l}dP^{tot}_{E}+\sum_{i}({\mu_{i}}dN_{i})
\end{equation}
\par

where
\par
\[
E_{l}=\frac{\partial{U}}{\partial{P^{tot}_{E}}}
\]
\par
$P^{tot}_{E}$ is the component of the electric moment parallel to
the external field.
\par
Thus, the system consisting of several chemical components in the
presence of an electrical filed takes the form
\par
\begin{equation}
\eta_{U ij}=\begin{pmatrix}
  \frac{T}{C_{v}} & -\frac{p}{C_{v}} & \frac{\partial{T}}{\partial{P^{tot}_{E}}} & \frac{\partial{T}}{\partial{N_{j}}} \\
  -\frac{p}{C_{v}} & \frac{pC_{p}}{vC_{v}} & -\frac{\partial{p}}{\partial{P_{E}}} & -\frac{\partial{p}}{\partial{N_{j}}} \\
  \frac{\partial{E_{l}}}{\partial{S}} & \frac{\partial{E_{l}}}{\partial{V}} & \frac{\partial{E}}{\partial{P^{tot}_{E}}} & \frac{\partial{E}}{\partial{N_{j}}} \\
  \frac{\partial{\mu_{i}}}{\partial{S}} & \frac{\partial{\mu_{i}}}{\partial{V}} & \frac{\partial{\mu_{i}}}{\partial{P^{tot}_{E}}} &
  \frac{\partial{\mu_{i}}}{\partial{N_{j}}}
\end{pmatrix}
\end{equation}
\par

\section{Curvature of TD-metric of a general thermodynamical system}
\par
In this section, we consider a thermodynamical potential
$E=E(x_{i})$ as function of the extensive variables $x_{i}$ and
calculate the curvature of the corresponding thermodynamical
metric defined on the integral (constitutive) surface
$\Sigma_{E}$, by the equation (2.3)

\[
\eta_{E ij}=\frac{\partial^{2}E}{\partial{x_{i}}\partial{x_j}}.
\]
\par
 Christoffel symbols for this metric are given by
\par
\begin{equation}
\Gamma^{k}_{ij}=\frac{1}{2}\sum_{m}\eta_{ij,m}\eta^{km},
\end{equation}
where $\eta_{ij,m}=\frac{\partial \eta_{ij}}{\partial x^{m}}.$
\par
After some calculations, it can be shown that the curvature tensor
of metric $\eta_{E}$ is given by
\par
\begin{equation}
R^{l}_{ijk}=
\Gamma^{l}_{ki,j}-\Gamma^{l}_{ji,k}+\Gamma^{l}_{jp}\Gamma^{p}_{ki}-\Gamma^{l}_{kp}\Gamma^{p}_{ji}=
\frac{1}{4}(\eta_{ij,m}\eta_{sn,k}-\eta_{sn,j}\eta_{ki,m})\eta^{mn}\eta^{ls}
\end{equation}
\par
Therefore, Ricci Tensor of metric $\eta_{E}$ is
\par
\begin{equation}
R_{ik}=R^{j}_{ijk}=\frac{1}{4}(\eta_{ij,m}\eta_{sn,k}-\eta_{sn,j}\eta_{ki,m})\eta^{mn}\eta^{js},
\end{equation}
and its scalar curvature $R(\eta )$ is, see also \cite{R3},
\begin{equation}
R=R_{ik}\eta^{ik}=\frac{1}{4}(\eta_{ij,m}\eta_{sn,k}-\eta_{sn,j}\eta_{ki,m})\eta^{mn}\eta^{js}\eta^{ik}.
\end{equation}
\par

\section{Hessian Surface of a thermodynamical potential and the sign of scalar curvature $R(\eta_{E})$}

In this section we study the Hessian representation of a metric
$\eta_{E}$ of a 5-dim thermodynamical system with two "degrees of
freedom" (using the terminology of Mechanics) $x^{i},\ i=1,2$ in
terms of an oriented 2D surface in 3D space of symmetrical
$2\times 2$ matrices.  We find a geometrical conditions on this
surface determining regions where scalar curvature of $\eta_{E}$
is positive, negative, zero, singular. In later sections we
demonstrate, on examples of ideal, van der Waals
 and Berthelot gases - to what physical consequences these
properties of metric $\eta_{U}$ lead to.\par

\subsection{Scalar curvature, 2D case}
For $k=2$ and a metric $\eta =\begin{pmatrix}\eta_{11} & \eta_{12}\\
\eta_{21} & \eta_{22}\end{pmatrix} $, we can use the known formula
for scalar curvature (\cite{Pog}, Ch.VIII, Sec.2)
\begin{equation}
R=-\frac{1}{4det(\eta )^{2}}det  \begin{pmatrix} \eta_{11} &
\eta_{11,1} & \eta_{11,2}\\
\eta_{12} &
\eta_{12,1} & \eta_{12,2}\\
\eta_{22} &
\eta_{22,1} & \eta_{22,2}\\
\end{pmatrix}  - \frac{1}{\sqrt{det(\eta)}}\left[  \left(
\frac{\eta_{11,2}-\eta_{12,1}}{\sqrt{det(\eta)}} \right)_{,2}-
\left( \frac{\eta_{12,2}-\eta_{22,1}}{\sqrt{det(\eta)}}
\right)_{,1}  \right] ,
\end{equation}
which, for the Weinhold metric reduces to the first term only:
\begin{equation}
R=-\frac{1}{4det(\eta )^{2}}det  \begin{pmatrix} \eta_{11} &
\eta_{11,1} & \eta_{11,2}\\
\eta_{12} &
\eta_{12,1} & \eta_{12,2}\\
\eta_{22} &
\eta_{22,1} & \eta_{22,2}\\
\end{pmatrix}.
\end{equation}
This expression shows that the scalar curvature goes to infinity
as $det(\eta) \rightarrow 0.$\par
\begin{remark}
Notice that the condition on a function $E$ that the metric
$\eta_{E}$ is {\bf degenerate everywhere}, i.e.
\[
det(\eta_{E})=\eta_{11}\eta_{22}-\eta_{12}^{2}=0\] is the simplest
{\bf Monge-Ampere equation},\cite{AG}. Geometry on the surfaces
defined by such a function is very interesting and well studied
(see, for instance \cite{AG} and the bibliography therein) but has
little relation to the case where $E$ is a thermodynamical
potential of a physically interesting system.
\end{remark}

\subsection{Hessian surface of a function $E$}
Scalar curvature $R$ is zero if and only if the determinant in the
numerator of expression (5.2) is zero. To determine where it can
take place, consider the {\bf Hessian mapping} $Hess_{E} $ of a
function $E$ mapping a 2D (canonically oriented) domain $D$ of
variables $x^{1},x^{2}$ into the 3D vector space of 2x2
symmetrical matrices $Sym(2,R)$ with the coordinates $\eta_{ij}$:
\begin{equation}
Hess_{E}:(x^{1},x^{2})\rightarrow \begin{pmatrix} E_{,11} &
E_{,12}\\ E_{,12} & E_{,22}
\end{pmatrix} .
\end{equation}
Image of this mapping is the 2D oriented surface $H_{E}$ in the
vector space $Sym(2,R).$ This surface is endowed with the frame
formed by the coordinate tangent vectors to this surface ${
\textbf{r}}_{i}(x)=Hess_{E\ *x}(\frac{\partial}{\partial x^{i}})$
at the nonsingular points of the surface where these vectors are
linearly independent. This determines orientation of the surface
$H_{E}$, or, what is the same, field of normal vectors $\textbf{
N}= \textbf{r}_{1}(x^{1},x^{2})\times
\textbf{r}_{2}(x^{1},x^{2}),$ where $\times $ is the standard
vector product in $R^3$.\par

Notice that, given a function $E$, the physically relevant domain
$D\subset R^2$ where mapping $Hess_{E}$ is defined can be smaller
then the natural domain $D(E)$ of the function $E$. As an example,
let us note the condition of {\bf positivity} of some physical
quantities (volume, pressure) or other lower boundary (Kelvin
temperature) condition.  Sometimes it is reasonable to assume that
function $E$ is defined on the boundary of domain $D$.
Correspondingly, not all the surface $H_{E}$ is interesting from
the point of view of physics.  Example of such a situation is
given below.\par

Condition of degeneracy of a matrix defines in the space
$Sym(2,R)$ the standard conic (nilpotent cone)
\begin{equation}
\eta_{11}\eta_{22}-\eta_{12}^{2}=0.
\end{equation}
Subset of degeneracy $sing(\eta_{E})\subset D$ of the metric
$\eta_{E}$ is the pre-image under the $Hess_{E}$ of the
intersection of surface $H_{E}$ and the cone (5.4).\par

\begin{example} As the first example of the $Hess(E)$ consider the ideal gas whose internal energy $U$ as a
function of variables $S,V$ and the mole number $N$ has the form
\[
U=U_{0}+C_{v}N\left(\frac{V}{N}\right) ^{-\frac{R}{C_{v}}}
e^{\frac{1}{C_{v}}(\frac{S}{N}-S_{0})},
\]
see \cite{KP}, Chapter 6. For one mole, taking $N=1$ we find
\begin{equation}
\eta_{U} =\frac{R}{C_{v}}e^{\frac{S}{C_{v}}} \begin{pmatrix}\frac{1}{R} V^{-\frac{R}{C_{v}}} & -V^{-\frac{C_{p}}{C_{v}}}\\
-V^{-\frac{C_{p}}{C_{v}}} & C_{p}V^{-(1+\frac{C_{p}}{C_{v}})}
\end{pmatrix} .
\end{equation}

\par Excluding variables $S,V$ from the formula for Weinhold metric, it is
easy to get equation for the surface $H_{U}$

\begin{equation}
R\eta_{11}\eta_{22}-C_{p}\eta_{12}^{2}=0.
\end{equation}
Thus the surface $\Sigma_{U}\subset Sym(2,R)$ is trivially
isomorphic to the standard {\bf conic} $xz-y^2=0$ in $R^{3}$.
Conic $(5.5)$ defines the homogeneous nonlinear second order
(Monge-Ampere) PDE
\[
RU_{,11}U_{,22}-C_{p}U_{,12}^{2}=0
\]
satisfied by the energy function $U$ of the ideal gas (see Theorem
2 below).\par

Comparing equations (5.4) and (5.6) we see that the signature
degeneration may happen on the Hessian surface of ideal gas only
in {\bf nonphysical} points $\eta_{12}=\eta_{11}\eta_{22}=0$ or
under the condition $C_{p}=R$, i.e. when $C_{v}=0$.
\end{example}
\begin{example}
As the second example, consider the van der Waals gas with molar
internal energy (9.6) and the Weinhold metric (9.8). Excluding
$e^{S/C_{v}}$ and $V$ from the equation (9.8) we get equation for
the Hessian surface $H_{U}$ for the vdW gas:
\begin{equation}
R\eta_{22}\eta_{11}-C_{p}\eta_{12}^{2}=-\frac{2aR\eta_{11}\eta_{12}^{3}}{(b\eta_{12}-R\eta_{11})^{3}}.
\end{equation}
Expression on the right is the (linear) correction to the ideal
gas equation for the Hessian surface ensuring nonzero curvature
almost everywhere on this surface.  Multiplying by the denominator
we get the algebraic equation for $H_{E}$
\[
(b\eta_{12}-R\eta_{11})^{3}(R\eta_{22}\eta_{11}-C_{p}\eta_{12}^{2})+2aR\eta_{11}\eta_{12}^{3}=0.
\]
given by the {\bf polynomial of the fifth order}.\par Intersection
of this surface with the surface $det(\eta_{ij})=0$ determines the
curve of the signature change (critical curve) for the vdW gas (in
this representation)
\[
\eta_{12}[2aR\eta_{11}\eta_{12}-C_{v}(b\eta_{12}-R\eta_{11})^{3}]=0
\]

\end{example}
\begin{remark}
It is interesting to notice that in these two examples the Hessian
surface $H_{E}$ is turned to be {\bf algebraic}. In general case
of $C_{v}$-const (see the constitutive equation (7.3)) this is not
so anymore.
\end{remark}

\vskip0.4cm
\subsection{Zero curvature condition}
 Equality of the determinant in the numerator of the formula for curvature
 to zero is equivalent to the linear dependency of radius vector $Hess_{E}(x)$
 of a point of the surface
and two tangent vectors $\textbf{r}_{i}(x)$. \par

Endow the space $Sym(2,R)$ with the canonical metric
\[
(A,B)=TR(AB),
\]
induced by the Killing metric on the Lie algebra $sl(2,R)$,
\cite{K}. Choosing  the orthonormal basis:
\[
X_{1}=\begin{pmatrix} 1 & 0 \\ 0 & 0 \end{pmatrix},\
X_{2}=\frac{1}{\sqrt{2}}\begin{pmatrix} 0 & 1 \\ 1 & 0
\end{pmatrix},\ X_{3}=\begin{pmatrix} 0 & 0 \\ 0 & 1
\end{pmatrix},\
\]
allows to identify $Sym(2,R)$ with $R^{3}$
\[
\begin{pmatrix} a & b \\ b & c \end{pmatrix} \longleftrightarrow
\begin{pmatrix} a \\ \sqrt{2}b \\ c \end{pmatrix}
\]
and to introduce scalar product with the standard Euclidian scalar
product $<\ ,\ > $.\par Then, the determinant in the numerator of
formula (5.2) takes the form
\begin{equation}
2^{3/2}<Hess_{E}(x^{1},x^{2}), \textbf{N}(x^{1},x^{2})>,
\end{equation}
where $\textbf{ N} $ is the normal vector to the surface $H_{E}$
introduced above. Thus, we can use alternatively picture of
surface $H_{E}$ in $Sym(2,R)$ and the standard Euclidian
$R^3$-representation of this surface and other objects. We will
use standard cartesian coordinates $(x,y,z)$ in $R^{3}$.
\par

Let $\Phi (x,y,z)=0$ be an equation of a surface $\Sigma ^{2}$ in
$R^{3}\equiv Sym(2,R)$ satisfying, on some open set $D\subset
\Sigma $ where the surface $\Sigma $ is smooth (nonsingular), to
the condition above: radius vector $\bar r$ at each point of the
surface is parallel (and belongs) to the tangent plane of the
surface at this point. This means, in particular, that tangent
planes at each point of the surface pass through the origin. Since
radius vector $\bar r$ at the point of the surface $\Phi
(x,y,z)=0$ belongs to the tangent plane of the surface at that
point, then
\[
\frac{\partial}{\partial t}\Phi ({\bar r}+t{\bar r})=0
\]
at each point of the domain $D$. From this it follows that
\[
{\bar r}\cdot \nabla \Phi =0
\]
on the subset $D$ of the surface $\Phi (x,y,z)=0$. In other words,
radial vector field  $q= {\bar r}\cdot \nabla $ is tangent to the
surface $\Phi (x,y,z)=0$.  Therefore, subset $D$ of this surface
is invariant under the (local) action of the group of dilations -
phase flow of vector field $q$.\par

 Thus, in an open subset $D$
where $R=0$, surface $H_{E}(D)$ is {\bf conical} and vice versa.
\par Consider now the projectivization $P_{2}(R)=P(Sym(2,R))$ of
the 3D vector space $Sym(2,R).$  Image of the conical surface
$H_{E}(D)$ in this projective space is one-dimensional and, near
nonsingular points (on an open subset of this image) is a smooth
curve. In homogeneous coordinates $(x,y,z)$ of $P(Sym(2,R))$, this
curve is (locally) given by an equation
\begin{equation}
L(x,y,z)=0,
\end{equation}
with a {\bf homogeneous function} $L$. This equation is also the
equation of the surface $H_{E}(D)\subset Sym(2,R)$ near almost all
points.\par
  Vice versa, if a surface in $Sym(2,R)$ is given by a
homogeneous equation of the type (5.5), it is conical. \par

As a result we see that components of Hessian of function $U$ must
satisfy to the following homogeneous (nonlinear in general) second
order PDE
\begin{equation}
L(E_{x^{1}x^{1}},E_{x^{1}x^{2}},E_{x^{2}x^{2}})=0.
\end{equation}
These arguments prove the following:
\begin{theorem} Let $k=2$, and let $E(x^{1},x^{2})$ be a smooth function. Following statements are equivalent
\begin{enumerate}
\item Weinhold metric $\eta_{E}$ has zero scalar curvature on an
open subset $D\subset R^{2}.$ \item Image of subset $D$ under the
Hessian mapping $Hess_{E}: R^{2}\rightarrow Sym_{2}(2,R)$ is a
{\bf conical} surface. \item Function $E$ satisfies to a {\bf
homogeneous} second order partial differential equation
\[
L(E_{x^{1}x^{1}},E_{x^{1}x^{2}},E_{x^{2}x^{2}})=0
\]
on the open set $D$.
\end{enumerate}
\end{theorem}
\par

\vskip0.4cm

 Consider in more details the case where the conical
surface $H_{E}$ is a plane. This case has no, as far as we know, a
direct physical application but is interesting as allowing to get
the explicit necessary and sufficient condition on the
thermodynamical potential $E$ to have zero scalar curvature
$R(\eta_{E)}).$\par

  In this case there is a
constant vector $\gamma \in Sym(2,R)$ such that radius vector
$Hess_{E}(x)$ of the surface is orthogonal to $\gamma $ at all
points $(x^{1},x^{2}) \in D$.\par In these terms, condition of
zero curvature above takes the form
\[
\gamma_{1}\eta_{11}(x)+\gamma_{2}\eta_{12}(x)+\gamma_{3}\eta_{22}(x)=0
\]
for all $x\in V$. Vector $\gamma $ is defined up to a
multiplication by a nonzero factor. \par Last equation can be
rewritten as the 2nd order linear PDE for the function
$U(x^{1},x^{2})$:
\begin{equation}
LU=\left (\gamma_{1}\frac{\partial ^{2}}{\partial x^{1\ 2}}+
\gamma_{2}\frac{\partial ^{2}}{\partial x^{1}x^{2}}+
\gamma_{3}\frac{\partial ^{2}}{\partial x^{2\ 2}} \right) U=0.
\end{equation}
Linear nondegenerate transformation in the plane $(x^{1},x^{2})$
transforms linear differential operator of the second order with
constant coefficients $L$ to one of three canonical forms:
\begin{enumerate}
\item If $\gamma_{1}\gamma_{3}-1/4\gamma_{2}^{2}>0$, to the (elliptic) form
\[
\Delta =\frac{\partial ^{2}}{\partial u^{1\ 2}}+\frac{\partial
^{2}}{\partial u^{2\ 2}}.
\]
\item
If $\gamma_{1}\gamma_{3}-1/4\gamma_{2}^{2}=0$, to the (degenerate)
form
\[
D=\frac{\partial ^{2}}{\partial u^{1\ 2}}+\frac{\partial
^{2}}{\partial u^{2\ 2}}.
\]
\item
If $\gamma_{1}\gamma_{3}-1/4\gamma_{2}^{2}<0$, to the (hyperbolic)
form
\[
\square =\frac{\partial ^{2}}{\partial u^{1\ 2}}-\frac{\partial
^{2}}{\partial u^{2\ 2}}.
\]
\end{enumerate}
Correspondingly, solutions of equation $(5.10)$ can be described
as the composition of an arbitrary solution of one of these three
model equations with a linear non-degenerate transformation in the
plane $(x^{1},x^{2})$.\par This finishes the proof of the first
part of the following theorem.
\begin{theorem}
For a thermodynamical system with two degrees of freedom and a
constitutive relation $E=E(x^{1},x^{2})$ the following statements
are equivalent:
\begin{enumerate}
\item Scalar curvature $R$ of the (Weinhold) metric $\eta
=\frac{\partial^{2}E}{\partial x^{i}\partial x^{j}}$ equals zero
in an open set $D\subset R^{2}$ and image of $D$ under the Hessian
mapping is the (part of) plane.

\item Each point $z\in D$ has a neighborhood $W$ such that the
generating function $E(x^{1},x^{2})$ satisfies to an equation
\begin{equation}
LE=\left (\gamma_{1}\frac{\partial ^{2}}{\partial x^{1\ 2}}+
\gamma_{2}\frac{\partial ^{2}}{\partial x^{1}x^{2}}+
\gamma_{3}\frac{\partial ^{2}}{\partial x^{2\ 2}} \right) E=0.
\end{equation}
with some  $\gamma \in R^{3}$ and
\begin{enumerate}
\item (elliptic case) If
$\gamma_{1}\gamma_{3}-1/4\gamma_{2}^{2}>0$, then $E$ is the
composition of a harmonic function on $R^{2}$ with a nondegenerate
linear transformation of the plane $(x^{1},x^{2})$. \item
(degenerate case) If $\gamma_{1}\gamma_{3}-1/4\gamma_{2}^{2}=0$,
then $E$ is the composition of an arbitrary function $f(x^{1})$ of
variable $x^{1}$ with a nondegenerate linear transformation of the
plane $(x^{1},x^{2})$. \item (hyperbolic case) If
$\gamma_{1}\gamma_{3}-1/4\gamma_{2}^{2}<0$, $E$ is the composition
of a function of the form $f(x^{1}+x^{2})+g(x^{1}-x^{2})$ with
arbitrary functions $f,g$ of one variable with a nondegenerate
linear transformation of the plane $(x^{1},x^{2})$.
\end{enumerate}
\item Statement of part two is valid in the whole open connected
set $D$ with the same vector $\gamma $.
\end{enumerate}
\end{theorem}
\begin{proof}
To prove the last statement of Theorem $2$, notice that, unless
surface $H_{E}$ degenerates to a straight line, $E$ cannot satisfy
to two equations with nonparallel vectors $\gamma ^{k}$. As a
result, local statement of the theorem can be extended to the
whole connected set $D$ with the same vector $\gamma $.
\end{proof}
\par
\subsection{Positive and negative scalar curvature - geometrical condition}
Generically, the set of points $(x^{1},x^{2})$ where
$R(\eta_{E})=0$ represent the curve (possibly singular i.e.
consisting of isolated points) on the plane $(x^{1},x^{2})$ -
pre-image under the mapping $Hess_{E}$ of the set - points that
are singular for the mapping $H_{E}\rightarrow P(Sym(2,R))$ of
projectivization.
\par

If at some point $(x^{1},x^{2})$, $R(x^{1},x^{2})\ne 0$, then the
sign of $R(x^{1},x^{2})$ is invariant under the relabelling of
variables $x^{1}$, namely, when we change $y^{1}=x^{2},\
y^{2}=x^{1}$, in the expression $(5.2)$ for curvature, denominator
is not changing while the two last columns of the determinant of
the matrix in the numerator are permuted. But at the same time,
first and last rows of the matrix in the numerator are permuted
too restoring the sign of the scalar curvature.\par

\begin{definition} Let $\Sigma^{2}$ be an oriented surface in $\R
^3$.  We call the surface $\Sigma^{2}$ (with the normal $\bar N$),
{\bf radially convex} (correspondingly, {\bf radially concave}))
if ${\bar r}\cdot {\bar N}>0$ for all ${\bar r}\in \Sigma^{2}$
(correspondingly ${\bar r}\cdot {\bar N}<0$ for all ${\bar r}\in
\Sigma^{2}$). \end{definition}

\par Then, as the arguments above
shows, the following statement is true,
\begin{theorem} Let $V\subset H_{E}$ be an open subset of the surface $H_{E}$. Following statements
 are equivalent:
\begin{enumerate}
\item $R>0$ (correspondingly $R<0$) for ${\bar r}\in V$,

\item Surface $H_{E}\vert_{V}$ is radially convex (correspondingly
radially concave).
\end{enumerate}
\end{theorem}
\vskip0.4cm
\subsection{Conformal equivalence of Weinhold and Ruppeiner metrics and the curvatures}
\par
\par
Recalling (\cite{C,M}) that thermodynamical metrics $\eta_{U}$ and
$\eta_{S}$ are related by contact transformations, generated by
the mapping $\phi$ of canonical "internal energy" chart $(U,
(T,S),(-P,V),(\mu_{i},N_{i}))$ to the entropy chart
 $(S,(-T^{-1},U),(\frac{-p}{T},V),(\frac{\mu_{i}}{T},N_{i}))$. More
specifically, we have, for the contact form in these two charts
\begin{equation}
\phi^{*}\omega _{S}=\tau \omega _{U},\ \tau =T^{-1},
\end{equation}
Correspondingly, equilibrium surface $\Sigma_{U}$ is mapped onto
the corresponding surface $\Sigma_{S}$. For {\bf metrics
$\eta_{U},\ \eta_{S}$ on these surfaces} we have the following
conformal equivalence relation (\cite{SNI,M2}:
\begin{equation}
 \phi^{*}\eta_{S} = \tau \eta_{U}, \ \tau =T^{-1},
\end{equation}
noticeably with the same conformal factor $\frac{1}{T}$ as the
contact form $\omega.$
\par
Recall (\cite{Po}, Ch.18) that if two metrics on a smooth manifold
$M$ are conformally equivalent:
\begin{equation}
{\bar g}=e^{2\sigma}g,
\end{equation}
with some function $\sigma \in C^{\infty}(M)$, then, their scalar
curvatures are related by
\begin{equation}
e^{2\sigma}R({\bar g})=R(g)-2(n-1)S,
\end{equation}
where $n$ is the dimension of the manifold $M$ and
$S=g^{ik}S_{ik}$, with
\[
S_{ij}=\nabla^{g}_{
j}\sigma_{i}-\sigma_{,i}\sigma_{,j}+\frac{1}{2}g_{ij}g^{kl}\sigma_{,k}\sigma_{,l}.
\]
Thus after conformal transformation one get in the expression for
scalar curvature the additional term
\begin{equation}
S= g^{ij}S_{ij}=g^{ij}\nabla^{g}_{
j}\sigma_{i}-g^{ij}\sigma_{,i}\sigma_{,j}+\frac{1}{2}g^{ij}g_{ij}g^{kl}\sigma_{,k}\sigma_{,l}=
\nabla^{g}_{ j}\sigma^{j}-\Vert d\sigma
\Vert_{g}^{2}+\frac{n}{2}\Vert d\sigma \Vert_{g}^{2}.
\end{equation}
First term on the right in the last part of this formula is the
covariant divergence with respect to the original metric $g$ of
the gradient (with respect to $g$) of function $\sigma$:
\[ div_{g}(\nabla^{g}\sigma )=\Delta_{g}\sigma .
\]
Thus,
\[
S=\Delta_{g}\sigma +\frac{n-2}{2}\Vert d\sigma \Vert_{g}^{2}.
\]
For $n=2$, the second term vanishes and we get for scalar
curvatures the following relation
\begin{equation}
e^{2\sigma}R({\bar g})=R(g)-2(\Delta_{g}\sigma ).
\end{equation}
In our case above $\sigma = -\frac{1}{2}ln(T)$, so we get relation
between curvatures of Ruppeiner $\eta_{S}$ and Weinhold $\eta_{U}$
(expressed in coordinates $(S,V)$):
\begin{equation}
R(\eta_{S})=TR(\eta_{U})+T\Delta_{g}ln(T).
\end{equation}
It follows from this that scalar curvatures of both metrics are
zero simultaneously if and only if the logarithm of temperature
$ln(T)$ is the {\bf harmonic function} of $(S,V)$.
\par

\section{Curvature of a general 2-D elementary system}
\par
\par
In this section we study the scalar curvature of a general 2-dim
elementary thermodynamical system. \par
 We found in Section 3 the Weinhold metric of any two-dimensional elementary system of entropy S and
volume V to be, $(3.14)$,(\cite{M2}),:
\[
\eta_{U} =(\eta_{U ij} )=\frac{1}{C_{v}}
\begin{pmatrix}
 T & -\frac{T\alpha}{k}\\
 -\frac{T\alpha}{k} & \frac{C_{p}}{vk}
\end{pmatrix}
\]

Determinant of the tensor $\eta_{U}$ (with respect to the standard
Euclidian metric) is

\begin{equation}
det(\eta_{U
ij})=\frac{TC_{p}}{VkC^{2}_{v}}-\frac{T^{2}\alpha^{2}}{k^{2}C^{2}_{v}}
=\frac{T}{kC^{2}_{v}}(\frac{C_{p}}{V}-\frac{T\alpha^{2}}{k})
=\frac{T}{kC^{2}_{v}}(\frac{C_{v}}{V}+\frac{T\alpha^{2}}{k}-\frac{T\alpha^{2}}{k})
=\frac{T}{kVC_{v}}
\end{equation}
\par
Using definition of k as above, we finally get
\par
\begin{equation}
det(\eta_{U ij})=-\frac{T}{C_{v}}(\frac{\partial p}{\partial
V})_{T}
\end{equation}
\par
Excluding T from being zero, the metric $\eta_{U}$ is degenerate
along the curve
\par
\begin{equation}
(\frac{\partial p}{\partial V})_{T}=0
\end{equation}
\par
which is usually presented in one of two forms: $p=p(V)$ or/and
$T=T(V)$. The critical triple point of the system is the extremum
point of these functions.\par
 Note that it is possible to write the isothermal
compressibility in terms of the determinant:
\par
\begin{equation}
k=\frac{T}{VC_{v}det(\eta_{U ij})}
\end{equation}
\par
\par
\begin{remark}
Determinant of our matrix $\eta_{U ij}$  is the denominator of the
expression (5.2) for the scalar curvature $R$ for such a system.
This implies that if $det(\eta_{U ij})\rightarrow{0}$, then
$R\rightarrow{\infty}.$ Therefore, along this curve the Weinhold
metric of the thermodynamical phase space changes signature and
its curvature is singular. As we will see, in the example of the
van der Waals gas, this is also the curve along which phase
transition occurs. Thus, this curve determines the boundary of the
equilibria region $\mathcal Eq$ on the constitutive surface
$\Sigma_{U}$.
\end{remark}
\par
\mathstrut
\par
\textbf{Remark: Speed of sound.}
\par
Let $\nu^{i}_{sound}$ and $\nu^{a}_{sound}$ respectively be the
speed of sound in terms of isothermal and adiabatic
compressibility $k$ and $k_{S}$. It is known that
$k_{S}=\frac{C_{v}}{C_{p}}k$. So, the speed of sound in a
thermodynamical system is given by
\par
\[
\nu^{a}_{sound}=\sqrt{\frac{vC_{p}det(\eta_{U
ij})}{T\rho}}\propto{\frac{1}{R}}
\]
\par
or
\par
\[
\nu^{i}_{sound}=\sqrt{\frac{vC_{v}det(\eta_{U
ij})}{T\rho}}\propto{\frac{1}{R}}
\]
\par
where R is the scalar curvature of the system.
\par
\mathstrut
\par
In region where $det(\eta_{U})\ne 0$ the inverse of $\eta_{U ij}$
is given by
\par
\begin{equation}
\eta^{ij}_{U}=\begin{pmatrix}
 \frac{C_{p}}{T} & V\alpha\\
 V\alpha & kV
\end{pmatrix}
\end{equation}
\par
Let's calculate now the third derivatives of the energy E. We get
\par
\begin{equation}
\eta_{{11,}_{1}}=\frac{T}{C^{2}_{v}}(1-(\frac{\partial
C_{v}}{\partial S})_{v})
\end{equation}
\par
\begin{equation}
\eta_{{11,}_{2}}=-\frac{T}{C^{2}_{v}}(\frac{\alpha}{k}+(\frac{\partial
C_{v}}{\partial V})_{s})
\end{equation}
\par
\begin{equation}
\eta_{{12,}_{2}}=\frac{1}{k^{2}C^{2}_{v}}(-kTC_{v}(\frac{\partial
\alpha}{\partial V})_{s}+Tk\alpha(\frac{\partial C_{v}}{\partial
V})_{s}+TC_{v}\alpha(\frac{\partial k}{\partial
V})_{s}+T\alpha^{2})
\end{equation}
\par
\[
\eta_{{22,}_{2}}=\frac{1}{k^{2}C^{2}_{v}}(2TC_{v}\alpha(\frac{\partial
\alpha}{\partial V})_{s}-T\alpha^{2}(\frac{\partial
C_{v}}{\partial
V})_{s}-\frac{T\alpha^{3}}{k}-\frac{kC^{2}_{v}}{V^{2}}
\]
\par
\begin{equation}
-(\frac{C^{2}_{v}}{V}+\frac{2TC_{v}\alpha^{2}}{k})(\frac{\partial
k}{\partial V})_{s})
\end{equation}
\par
Now, since $\eta_{{11,}_{2}}=\eta_{{12,}_{1}}$, where
\par
\begin{equation}
\eta_{{12,}_{1}}=-\frac{T}{C^{2}_{v}}(\frac{\alpha}{k}(1-(\frac{\partial
C_{v}}{\partial S})_{v})+\frac{C_{v}}{k}(\frac{\partial
\alpha}{\partial S})_{v}-\frac{C_{v}\alpha}{k^{2}}(\frac{\partial
k}{\partial S})_{v})
\end{equation}
\par
we get the following
\par
\begin{lemma}
Identity I.
\begin{equation}
(\frac{\partial{C_{v}}}{\partial{V}})_{s}+\frac{\alpha}{k}(\frac{\partial{
C_{v}}}{\partial {S}})_{v}=\frac{C_{v}}{k}(\frac{\partial{
\alpha}}{\partial{
S}})_{v}-\frac{C_{v}\alpha}{k^{2}}(\frac{\partial{k}}{\partial{
S}})_{v}
\end{equation}
\end{lemma}
\par
Similarly, since $\eta_{{12,}_{2}}=\eta_{{22,}_{1}}$, where
\par
\begin{equation}
\eta_{{22,}_{1}}=\frac{1}{k^{2}C^{2}_{v}}(T\alpha^{2}(1-(\frac{\partial
C_{v}}{\partial S})_{v}+2TC_{v}\alpha(\frac{\partial
\alpha}{\partial
S})_{v}-\frac{2TC_{v}\alpha^{2}}{k}(\frac{\partial k}{\partial
S})_{v} -\frac{C^{2}_{v}}{V}(\frac{\partial k}{\partial S})_{v})
\end{equation}
\par
we have
\par
\begin{lemma}
Identity II
\begin{equation}
k(\frac{\partial}{\partial{V}}\ln{\frac{k}{\alpha}})_{S}=(\frac{\partial\alpha}{\partial
S})_{v}-(\frac{\alpha}{k}+\frac{C_{v}}{TV\alpha})(\frac{\partial
k}{\partial S})_{v}
\end{equation}
\end{lemma}
\par
Now, for the components of the Ricci Curvature we have
\par
\begin{equation}
R_{11}=\frac{1}{4}(((\eta_{{21,}_{1}})^{2}-\eta_{{11,}_{1}}\eta_{{21,}_{2}})\eta^{11}\eta^{22}+(\eta_{{21,}_{1}}\eta_{{21,}_{2}}-\eta_{{11,}_{1}}\eta_{{22,}_{2}})\eta^{12}\eta^{22}+((\eta_{{21,}_{2}})^{2}-\eta_{{11,}_{2}}\eta_{{22,}_{2}})(\eta^{22})^{2})
\end{equation}
\par
\begin{equation}
R_{12}=R_{21}=\frac{1}{4}((\eta_{{11,}_{1}}\eta_{{21,}_{2}}-(\eta_{{21,}_{1}})^{2})\eta^{11}\eta^{12}+(\eta_{{11,}_{1}}\eta_{{22,}_{2}}-\eta_{{21,}_{2}}\eta_{{11,}_{2}})({\eta^{12}})^{2}+(\eta_{{11,}_{2}}\eta_{{22,}_{2}}-(\eta_{{21,}_{2}})^{2})\eta^{12}\eta^{22})
\end{equation}
\par
\begin{equation}
R_{22}=\frac{1}{4}((\eta_{{21,}_{1}})^{2}-\eta_{{22,}_{1}}\eta_{{11,}_{1}})(\eta^{11})^{2}+(\eta_{{21,}_{2}}\eta_{{11,}_{2}}-\eta_{{11,}_{1}}\eta_{{22,}_{2}})\eta^{11}\eta^{12}+((\eta_{{21,}_{2}})^{2}-\eta_{{12,}_{1}}\eta_{{22,}_{2}})\eta^{11}\eta^{22})
\end{equation}
\par
It is easy to show that
\par
\begin{equation}
R_{11}\eta^{11}=R_{22}\eta^{22}
\end{equation}
\par
and so, scalar curvature is given by,
\par
\begin{equation}
R=2(R_{11}\eta^{11}+R_{12}\eta^{12})
\end{equation}
\par
After some calculation, we get
\par
\begin{equation}
R_{11}=\frac{T^{2}}{4C^{4}_{v}det(\eta_{ij})}[HG+\frac{C_{v}\alpha}{k^{2}}F(\frac{TV\alpha}{k}F-J)]
\end{equation}
\par
where
\par
\begin{equation}
H=(\frac{\alpha}{k}-\frac{C_{v}}{V}+\frac{C_{p}-C_{v}}{\alpha}(\frac{\partial
\alpha}{\partial V})_{s}-\frac{C_{p}}{k}(\frac{\partial
k}{\partial V})_{s}+(\frac{\partial C_{v}}{\partial V})_{s})
\end{equation}
\par
\begin{equation}
G=(\frac{\partial C_{v}}{\partial
V})_{s}+\frac{\alpha}{k}(\frac{\partial C_{v}}{\partial S})_{v}
\end{equation}
\par
\begin{equation}
F=k(\frac{\partial}{\partial{V}}\ln{\frac{k}{\alpha}})_{S}
\end{equation}
\par
and
\par
\begin{equation}
J=1-(\frac{\partial{C_{v}}}{\partial{S}})_{V}
\end{equation}
\par
Then,from the $(1,1)$ component of the Ricci Curvature we derive
that
\par
\begin{equation}
R_{12}=R_{21}=-\frac{\alpha}{k}R_{11}
\end{equation}
\par
and
\par
\begin{equation}
R_{22}=\frac{C_{p}}{TVk}R_{11}
\end{equation}
\par
Therefore, Ricci tensor have the form
\par
\begin{equation}
R_{ij}=R_{11}
\begin{pmatrix}
 1 & -\frac{\alpha}{k}\\
 -\frac{\alpha}{k} & \frac{C_{p}}{TVk}
\end{pmatrix}
=\frac{C_{v}}{T}\eta_{ij}R_{11}
\end{equation}
\par
Scalar curvature is equal to
\begin{equation}
R=\frac{2C_{v}}{T}R_{11}
\end{equation}
 and using (6.19) we get the following result,
\par
\begin{theorem}
Scalar curvature of a general two-dimensional elementary system is
given by
\par
\begin{equation}
R=\frac{T}{2C^{3}_{v}det(\eta_{ij})}[HG+\frac{C_{v}\alpha}{k^{2}}F(\frac{TV\alpha}{k}F-J)]
\end{equation}
\par
where
\par
\begin{equation}
H=(\frac{\alpha}{k}-\frac{C_{v}}{V}+\frac{C_{p}-C_{v}}{\alpha}(\frac{\partial
\alpha}{\partial V})_{s}-\frac{C_{p}}{k}(\frac{\partial
k}{\partial V})_{s}+(\frac{\partial C_{v}}{\partial V})_{s}),
\end{equation}
\par
\begin{equation}
G=(\frac{\partial C_{v}}{\partial
V})_{s}+\frac{\alpha}{k}(\frac{\partial C_{v}}{\partial S})_{v},\
\ F=k(\frac{\partial}{\partial{V}}\ln{\frac{k}{\alpha}})_{S}
\end{equation}
\par
and
\par
\begin{equation}
J=1-(\frac{\partial{C_{v}}}{\partial{S}})_{v}
\end{equation}
\par
\end{theorem}
\par
\par
\section{Case of constant $C_{v}$}
\par

Here we discuss the class of systems where $C_{v}$ is constant -
class where both the ideal gas and van der Waals gas are included.
In this case we can easily get the general form of the
constitutive law $U=U(S,V)$, calculations of the previous section
simplifies and we will be able to discuss curvature of metric
$\eta_{U}$ in more details.\par
\begin{remark}
Notice that if molar internal energy $U$ is given as a function of
volume $V$ and the temperature $T$, then the Helmholz equation
\[
(\frac{\partial U}{\partial V})_{T}=T^{2}\frac{\partial}{\partial
T}\left( \frac{p}{T}\right)\vert_{V},
\]
leads (see \cite{KP}, Ch. 6) to the formula
\[
C_{v}=C_{v_{ ideal}}+\int_{\infty}^{V}T\left( \frac{\partial
^{2}p}{\partial T^{2}} \right)_{V}dV.
\]
relating the heat capacity of any 2TD-system to one for ideal or
vdW gas. From this formula it follows that, for a 2TD-system for
which the integral above converges, heat capacity $C_{v}$ is
constant if and only if
\[
\left( \frac{\partial ^{2}p}{\partial T^{2}} \right)_{V}=0,
\]
(i.e. when $p$ is the {\bf linear} function of $T$).\par
\end{remark}

Let us find possible constitutive laws $S=S(T,V),\ U=U(S,V)$ for
which $C_{v}-const.$ Recall that $C_{v}=T(\frac{\partial
S}{\partial T})_{V}$. Rewrite this as
\[
\frac{\partial S}{\partial T}=\frac{C_{v}}{T},
\]
and integrate:
\begin{equation}
S=C_{v}ln(T)+f(V),
\end{equation}
with arbitrary function $f(V)$.\par

On the other hand, $T=\frac{\partial U}{\partial S}$. Substituting
this in the definition of $C_{v}$ in the form $\frac{\partial
T}{\partial S}=\frac{T}{C_{v}}$, we get for $U$ equation
\begin{equation}
\frac{\partial ^{2}U}{\partial
S^{2}}-\frac{1}{C_{v}}\frac{\partial U}{\partial S}=0.
\end{equation}
Integrating once we get
\[
\frac{\partial U}{\partial S}=\frac{U}{C_{v}}+g(V),
\]
with an arbitrary function $g(V)$. Setting $f=f_{1}$ and $g=f_{2}$
and solving this equation we get the fundamental constitutive law
in the form
\begin{equation}
U(S,V)=f_{1}(V)e^{S/C_{v}}-C_{v}f_{2}(V),
\end{equation}
with arbitrary functions $f_{i}(V)$.\par
\begin{example}
Case of the ideal gas is obtained here if we take
$f_{1}=V^{-\frac{R}{C_{v}}},\ f_{2}=0$ since
$U=V^{-\frac{R}{C_{v}}}e^{\frac{S}{C_{v}}}$ with $U_{0}=0$ and
$S_{0}=0$ .(see section $9$).
\end{example}
\par
\begin{example}
For the Van der Waals gas we take
$f_{1}=(V-b)^{-\frac{R}{C_{v}}},\ f_{2}=\frac{a}{C_{v}V}$ since
$U=(V-b)^{-\frac{R}{C_{v}}}e^{\frac{S}{C_{v}}}-\frac{a}{V}$ with
$U_{0}=0$ and $S_{0}=0$.(see section 10).
\end{example}
\par

Taking derivatives by both variables and excluding $S$ from these
relations we find the following state equation
\begin{equation}
p=-C_{v}T\frac{f_{1}'(V)}{f_{1}(V)}+C_{v}f'_{2}(V).
\end{equation}
Calculating derivative of (7.3) and denoting $f_{i}(V)=f_{i}$ for
i=1,2, we get
\par
\begin{equation}
(\frac{\partial U}{\partial
S})_{v}=\frac{f_{1}}{C_{v}}e^\frac{S}{C_{v}},\ (\frac{\partial
U}{\partial V})_{s}=f^{'}_{1}e^\frac{S}{C_{v}}-C_{v}f^{'}_{2},
\end{equation}
\par
\begin{equation}
\eta_{11}=(\frac{\partial^{2} U}{\partial
S^{2}})_{v}=\frac{f_{1}}{C^{2}_{v}}e^\frac{S}{C_{v}}
\end{equation}
\par
\begin{equation}
\eta_{12}=\eta_{21}=(\frac{\partial^{2} U}{\partial S \partial
V})=\frac{f^{'}_{1}}{C_{v}}e^\frac{S}{C_{v}}
\end{equation}
\par
\begin{equation}
\eta_{22}=(\frac{\partial^{2} U}{\partial
V^{2}})_{s}=f^{''}_{1}e^\frac{S}{C_{v}}-C_{v}f^{''}_{2}
\end{equation}
As a result we get the metric $\eta_{U}$ in the form

\par
\begin{equation}
\eta_{U ij}=\begin{pmatrix}
  \frac{f_{1}}{C^{2}_{v}}e^\frac{S}{C_{v}} & \frac{f^{'}_{1}}{C_{v}}e^\frac{S}{C_{v}} \\
  \frac{f^{'}_{1}}{C_{v}}e^\frac{S}{C_{v}} & f^{''}_{1}e^\frac{S}{C_{v}}-C_{v}f^{''}_{2} \\
\end{pmatrix}
\end{equation}
\par
with the determinant
\par
\[
\det\eta_{ij}=\frac{e^\frac{2S}{C_{v}}}{C^{2}_{v}}(f_{1}f^{''}_{1}-(f^{'}_{1})^{2}-e^\frac{-S}{C_{v}}C_{v}f_{1}f^{''}_{2})
\]
\par
\begin{equation}
=\frac{e^\frac{2S}{C_{v}}}{C^{2}_{v}}(f_{1}f^{''}_{1}-(f^{'}_{1})^{2})-\frac{e^\frac{S}{C_{v}}}{C^{2}_{v}}f_{1}f^{''}_{2}
\end{equation}
\par
Consider, now, a special case of a constitutive relation (7.9)
with the same $f_{1}$ and with $f_{2}=0$(ideal gas), and mark this
case with $0$. For this case
\par
\begin{equation}
\det\eta_{{ij,}_{0}}=\frac{e^\frac{2S}{C_{v}}}{C^{2}_{v}}(f_{1}f^{''}_{1}-(f^{'}_{1})^{2})
\end{equation}
\par
This determinant is zero if and only if
$f_{1}f^{''}_{1}-(f^{'}_{1})^{2}=0$, i.e. if
$f_{1}=c_{1}e^{c_{2}V}$ with positive constants $c  _{i}$.
\par
Now we have the important relation
\par
\begin{lemma}
\begin{equation}
\det\eta_{{ij,}_{real}}=\det\eta_{{ij,}_{0}}-\frac{e^\frac{S}{C_{v}}}{C^{2}_{v}}f_{1}f^{''}_{2}
\end{equation}
\end{lemma}
\par

Calculate third derivatives of $U$ of the form (6.3):
\par
\begin{equation}
\eta_{{11,}_{1}}=\frac{f_{1}}{C^{3}_{v}}e^\frac{S}{C_{v}}=\frac{1}{C_{v}}\eta_{11}
\end{equation}
\par
\begin{equation}
\eta_{{12,}_{1}}=\eta_{{21,}_{1}}=\eta_{{11,}_{2}}=\frac{f^{'}_{1}}{C^{2}_{v}}e^\frac{S}{C_{v}}=\frac{1}{C_{v}}\eta_{12}
\end{equation}
\par
\begin{equation}
\eta_{{12,}_{2}}=\eta_{{21,}_{2}}=\eta_{{22,}_{1}}=\frac{f^{''}_{1}}{C_{v}}e^\frac{S}{C_{v}}=\frac{1}{C_{v}}\eta_{22}+f^{''}_{2}
\end{equation}
\par
\begin{equation}
\eta_{{22,}_{2}}=f^{'''}_{1}e^\frac{S}{C_{v}}-C_{v}f^{'''}_{2}
\end{equation}
\par
Using these derivatives in formula (6.19) for the component
$R_{11}$ of the Ricci tensor we get:
\par
\begin{equation}
R_{11}=\frac{e^\frac{S}{C_{v}}f_{1}f^{''}_{2}}{4C_{v}e^\frac{S}{C_{v}}(f_{1}f^{''}_{1}-(f^{'}_{1})^{2}-e^\frac{-S}{C_{v}}C_{v}f_{1}f^{''}_{2})^{2}}(f_{1}f^{''}_{1}-(f^{'}_{1})^{2})
\end{equation}
\par
which can be written as
\par
\begin{equation}
R_{11}=\frac{e^\frac{S}{C_{v}}f_{1}f^{''}_{2}}{4C^{3}_{v}}\frac{\det\eta_{{ij,}_{0}}}{(\det\eta_{{ij,}_{real}})^{2}}
\end{equation}
\par
So, finally we get scalar curvature to be
\par
\begin{equation}
R=\frac{e^\frac{S}{C_{v}}f_{1}f^{''}_{2}}{2TC^{2}_{v}}\frac{\det\eta_{{ij,}_{0}}}{(\det\eta_{{ij,}_{real}})^{2}}=\frac{Tf_{1}f^{''}_{2}[f_{1}f^{''}_{1}-(f^{'}_{1})^{2}]}{2[T(f_{1}f^{''}_{1}-(f^{'}_{1})^{2})-f^{2}_{1}f^{''}_{2}]^{2}}
\end{equation}
\par
This leads to the following statement:
\begin{proposition}
Let $C_{v}=const$ and let the internal energy $U(S,V)$ be given by
the fundamental constitutive relation
\begin{equation}
U(S,V)=f_{1}(V)e^{S/C_{v}}-C_{v}f_{2}(V),
\end{equation}
then the curvature of Weinhold metric is zero in a domain
$D\subset R^{2}$,
\[
R(\eta_{U})\vert_{D}=0
\]
if and only if one of three cases holds:
\begin{enumerate}
\item
\[
f_{1}=c_{1}e^{c_{2}V},
\]
for some positive constants $c_{i}$,
\item
\[
f_{2}=AV+B,
\]
with some constants $A,B$, that includes the case of ideal gas.
\item
\[
f_{1}(V)=0,
\]
degenerate case.
\end{enumerate}
\end{proposition}

 It is interesting to see how the scalar curvature is expressed, in the case $C_{v}-const$, in
 terms of isothermal compressibility $k$.  We calculate
\par
\begin{equation}
\eta_{{11,}_{1}}=\frac{T}{C^{2}_{v}}=\frac{1}{C_{v}}\eta_{11},\
\eta_{{11,}_{2}}=\eta_{12,1}=-\frac{T\alpha}{C^{2}_{v}k}=\frac{1}{C_{v}}\eta_{12},
\end{equation}
\par
\begin{equation}
\eta_{{12,}_{2}}=\eta_{22,1}=\frac{1}{k^{2}C^{2}_{v}}(\alpha^{2}T-\frac{C^{2}_{v}}{V}(\frac{\partial
k}{\partial
S})_{v})=\frac{1}{T}((\eta_{12})^{2}-\frac{T}{Vk^{2}}(\frac{\partial
k}{\partial S})_{v}),
\end{equation}
\par
\begin{equation}
\eta_{{22,}_{2}}=\frac{1}{k^{2}C^{2}_{v}}(2\frac{C^{2}_{v}\alpha}{Vk}(\frac{\partial
k}{\partial S})_{v}-\frac{C^{2}_{v}}{V}(\frac{\partial k}{\partial
V})_{s}-\frac{T\alpha^{3}}{k}-\frac{kC^{2}_{v}}{V^{2}}).
\end{equation}
\par

Moreover $(6.15)$ reduces to the following
\par
\begin{lemma}
Identity III
\begin{equation}
(\frac{\partial \alpha}{\partial k})_{v}=\frac{\alpha}{k}
\end{equation}
\end{lemma}
\par
From which we also get that $(6.13)$ becomes
\begin{equation}
\alpha(\frac{\partial}{\partial{V}}\ln{\frac{\alpha}{k}})_{S}=\frac{C_{v}}{TVk}(\frac{\partial
k}{\partial S})_{v}
\end{equation}
\par
\begin{proposition}
Scalar curvature of metric $\eta_{U}$, in the case of a constant
$C_{v}$, is given by
\begin{equation}
R=\frac{C_{v}}{2Tk^{2}}(\frac{\partial k}{\partial
S})_{v}((\frac{\partial k}{\partial
S})_{v}+\frac{k}{C_{v}})=\frac{C_{v}}{2T}\left( \frac{\partial
ln(k)}{\partial S}_{v} \right) \times \left( \frac{\partial
ln(k)}{\partial S}_{v}+\frac{1}{C_{v}} \right)
\end{equation}
\end{proposition}
\par
\begin{proof} Since $C_{v}$ is constant, we can use $(6.22)$ and $(7.25)$
to get
\par
\begin{equation}
F=-\frac{C_{v}}{TV\alpha}(\frac{\partial k}{\partial S})_{v}
\end{equation}
\par
Therefore, after some calculation and considering $G=0$ and
$det(\eta_{ij})=\frac{T}{kVC_{v}}$ , we get $(7.26).$
\end{proof}
\par
\begin{corollary} Curvature $R(\eta_{U})$ is {\bf negative} if and
only if
\[
-\frac{1}{C_{v}}<\frac{\partial ln(k)}{\partial S}\vert_{v}<0.
\]
\end{corollary}

\par
\section{Ideal Gas}
\par
Here we consider the ideal gas, the simplest model example of a TD
system with two degrees of freedom. For more detailed discussion
of Weinhold metric(s) for ideal gas in different representations
we refer to \cite{NS}. Except its illustrative interest, ideal gas
will serve us as a reference point in the further study of
curvature of metric $\eta_{E}$ for non-ideal gases.\par

 Internal energy $U$ of the
ideal gas as a function of $S,V$ and the mole number $N$ has the
form
\begin{equation}
U=U_{0}+C_{V}N\left(\frac{V}{N}\right) ^{-\frac{R}{C_{v}}}
e^{\frac{1}{C_{V}}(\frac{S}{N}-S_{0})},
\end{equation}
with some reference constants $U_{0},S_{0}$, see \cite{KP},
Chapter 6. Other equations of state are obtained from (8.1) in a
standard way:

\begin{equation}
pV=NRT,\ U=C_{V}NT,\ TS+\mu N=C_{P}NT.
\end{equation}
\par

We will take $N=1$. Then it is known that
\[
S=S_{0}+C_{v}\ln{U}+R\ln{V}
\]
\par
from which we can solve for $U$ obtaining
\par
\[
U=U(S,V)=U_{0}+C_{v}V^{-\frac{R}{C_{v}}}e^{\frac{S-S_{0}}{C_{v}}}
\]
\par

Weinhold metric for an Ideal Gas is given by (3.14). Setting
$S_{0}=0$, we get
\par
\begin{equation}
\eta_{U_{ij}} =\frac{1}{C_{v}}
\begin{pmatrix}
 T & -p\\
 -p & \frac{C_{p}p}{V} \end{pmatrix} =\frac{R}{C_{v}}e^{\frac{S}{C_{v}}} \begin{pmatrix}\frac{1}{R} V^{-\frac{R}{C_{v}}} & -V^{-\frac{C_{p}}{C_{v}}}\\
-V^{-\frac{C_{p}}{C_{v}}} & C_{p}V^{-(1+\frac{C_{p}}{C_{V}})}
\end{pmatrix} .
\end{equation}

\par
 In this case
$C_{v}$ and $C_{p}$ are positive constants. From $(5.5)$ we get
\par
\begin{equation}
det(\eta)=\frac{p}{VC^{2}_{v}}(TC_{p}-pV)=\frac{pT}{VC_{v}}=\frac{R}{C_{v}}e^{2\frac{S}{C_{v}}}V^{-2(\frac{C_{p}}{C_{v}})}>0
\end{equation}
\par
which is always positive.  Thus, Weinhold metric is {\bf positive
definite} on the constitutive surface $\Sigma_{U}$.
\par
Inverse of $\eta_{U ij}$ is given by
\par
\begin{equation}
\eta^{ij}_{U}=\begin{pmatrix}
 \frac{C_{p}}{T} & \frac{V}{T}\\
 \frac{V}{T} & \frac{V}{p}
\end{pmatrix}
\end{equation}
\par
Now, $(6.6)$ through $(6.9)$ become
\par
\begin{equation}
\eta_{{11,}_{1}}=\frac{T}{C^{2}_{v}},\qquad
\eta_{{11,}_{2}}=\eta_{{12,}_{1}}=\eta_{{21,}_{1}}=-\frac{p}{C^{2}_{v}}
\end{equation}
\par
\begin{equation}
\eta_{{22,}_{1}}=\eta_{{12,}_{2}}=\eta_{{21,}_{2}}=\frac{p\gamma}{vC_{v}},\qquad
\eta_{{22,}_{2}}=-(\gamma+1)\frac{p\gamma}{v^{2}}
\end{equation}
where $\gamma=\frac{C_{p}}{C_{v}}.$
\par
For the component $R_{11}$ of the Ricci Tensor $R_{ij}$ of ideal
gas we get
\par
\begin{equation}
R_{11}=(R-C_{p})\frac{C_{p}}{C^{4}_{v}}+(C_{p}+C_{v}-R)\frac{C_{p}}{C^{4}_{v}}-\frac{C_{p}}{C^{3}_{v}}=0.
\end{equation}
\par
Using (6.26), we get the known result (see \cite{NS}):
\par
\begin{equation}
R_{ij}=0 \qquad\ \forall i,j=1,2.
\end{equation}
\par
and, therefore,
\par
\begin{equation}
R(\eta_{U})=0
\end{equation}
\par
Zero curvature of Weinhold metric (as well as of Ruppeiner metric
based on entropy \cite{R2}) is positively correlated with the
known suggestion to use thermodynamical curvature as the
characteristic of interactions on the microscopic level of media
description.\par

\par

\mathstrut

\section{Van der Waals Gas}
Let's consider now a Van der Waals gas whose equation of state is
given by
\par
\begin{equation}
(p+\frac{a}{V^{2}})(V-b)=RT,
\end{equation}
where a and b are positive constants. This equation represents the
behavior of real gases more accurately than the ideal gas by
introducing two additional positive constants a and b
characteristic (of molecule interaction in the gas and of the part
of the volume occupied by the molecules correspondingly, see
\cite{C}) of the particular gas under consideration.
\par
For the coefficients $\alpha$ and $k$ we get
\par
\begin{equation}
\alpha=\frac{RV^{2}}{pV^{3}-aV+2ab}=\frac{RV^{2}(V-b)}{RTV^{3}-2a(V-b)^{2}}
\end{equation}
\par
and
\par
\begin{equation}
k=\frac{(V-b)V^{2}}{pV^{3}-aV+2ab}=\frac{V^{2}(V-b)^{2}}{RTV^{3}-2a(V-b)^{2}}
\end{equation}
\par
Therefore the relation between the coefficient of expansion and
the isothermal compressibility is given by
$\alpha=\frac{R}{(V-b)}k.$
\par
The entropy of VdW gas per one mole ($N=1$) is given by,\cite{KP},
\par
\begin{equation}
S=R\ln{(V-b)(U+\frac{a}{V})}^{\frac{C_{v}}{R}}+S_{0}.
\end{equation}
\par
The internal energy of Van der Waals gas as a function of S and V
is obtained from the last equation, namely
\par
\begin{equation}
U=U_{0}+(V-b)^{-\frac{R}{C_{v}}}e^{\frac{S}{C_{v}}}-\frac{a}{V}
\end{equation}
\par
Then, the Weinhold metric of this gas is given by
\par
\begin{equation}
\eta_{U ij}=\begin{pmatrix}
 \frac{T}{C_{v}} & -\frac{TR}{(V-b)C_{v}}\\
 -\frac{TR}{(V-b)C_{v}} & (\frac{TR}{(V-b)^{2}}(1+\frac{R}{C_{v}})-\frac{2a}{V^{3}})
\end{pmatrix}
\end{equation}
\par
\begin{equation}
=\begin{pmatrix}
\frac{1}{C^{2}_{v}}e^{\frac{S}{C_{v}}}(V-b)^{-\frac{R}{C_{v}}} &
-\frac{R}{C^{2}_{v}}e^{\frac{S}{C_{v}}}(V-b)^{-(1+\frac{R}{C_{v}})}\\
-\frac{R}{C^{2}_{v}}e^{\frac{S}{C_{v}}}(V-b)^{-(1+\frac{R}{C_{v}})}
&
\frac{R}{C^{2}_{v}}e^{\frac{S}{C_{v}}}(C_{v}+R)(V-b)^{-(2+\frac{R}{C_{v}})}-\frac{2a}{V^{3}}
\end{pmatrix},
\end{equation}
with $C_{v}$ being constant. In the limit $a\rightarrow 0,\
b\rightarrow 0$ to the Ideal Gas we get exactly the metric
$(8.3)$. The determinant is then given by
\par
\begin{equation}
det(\eta_{ij})=\frac{RT^{2}V^{3}-2aT(V-b)^{2}}{V^{3}C_{v}(V-b)^2}=\frac{T}{V^{3}(V-b)C_{v}}(pV^{3}-aV+2ab)
\end{equation}
\par
\begin{equation}
=\frac{1}{C^{2}_{v}}\frac{e^{\frac{S}{C_{v}}}}{(V-b)^{\frac{R}{C_{v}}}}[\frac{Re^{\frac{S}{C_{v}}}}{C_{v}(V-b)^{2+\frac{R}{C_{v}}}}-\frac{2a}{V^{3}}]
\end{equation}
\par
which is zero along the curve $\gamma_{\eta}$
\par
\begin{equation}
S=S(V)=C_{v}[(2+\frac{R}{C_{v}})\ln{(V-b)}+\ln{(\frac{2aC_{v}}{RV^{3}})}]
\end{equation}
\par
or, when $pV^{3}-aV+2ab=0$, supposing of course that our system is
at non-zero temperature.
\par
The determinant of the metric $\eta_{U}$ is positive if entropy is
large enough
\[
S=S(V)>C_{v}[(2+\frac{R}{C_{v}})\ln{(V-b)}+\ln{(\frac{2aC_{v}}{RV^{3}})}]
\]
and negative if the opposite is true.
\par
The inverse of tensor $\eta_{U ij}$ is given by
\par
\begin{equation}
\eta^{ij}_{U}=\begin{pmatrix}
 (\frac{C_{v}}{T}+\frac{R^{2}V^{3}}{RTV^{3}-2a(V-b)^{2}}) & \frac{RV^{3}(V-b)}{RTV^{3}-2a(V-b)^{2}}\\
 \frac{RV^{3}(V-b)}{RTV^{3}-2a(V-b)^{2}} & \frac{V^{3}(V-b)^{2}}{RTV^{3}-2a(V-b)^{2}}
\end{pmatrix}
\end{equation}.
\par
Now, we calculate the third derivatives of the energy U. We get
\par
\begin{equation}
\eta_{{11,}_{1}}=\frac{T}{C^{2}_{v}};\
\eta_{{12,}_{1}}=\eta_{{21,}_{1}}=\eta_{{11,}_{2}}=-\frac{RT}{(V-b)C^{2}_{v}};
\end{equation}
\par
\begin{equation}
\eta_{{12,}_{2}}=\eta_{{21,}_{2}}=\eta_{{22,}_{1}}=\frac{RT}{C_{v}(V-b)^{2}}(1+\frac{R}{C_{v}})
;\
\eta_{{22,}_{2}}=\frac{6a}{V^{4}}-\frac{RT}{(V-b)^3}(1+\frac{R}{C_{v}})(2+\frac{R}{C_{v}})
\end{equation}
\par
Then, the component $R_{11}$ of the Ricci Tensor is given by
\par
\begin{equation}
R_{11}=\frac{aRTV^{3}(V-b)^{2}}{2C^{2}_{v}(RTV^{3}-2a(V-b)^{2})^{2}}
\end{equation}
\par
From $(6.27)$, we get the scalar curvature $R(\eta_{U})$ of the
Weinhold metric to be
\par
\begin{equation}
R(\eta_{U})=\frac{aRV^{3}}{C_{v}(pV^{3}-aV+2ab)^{2}}
\end{equation}
\par
The scalar curvature goes to zero as $a\rightarrow{0}$ or as
$V\rightarrow{\infty}$. Since the quantity $\frac{a}{V^{2}}$
characterizes the attractive interaction within a system, scalar
curvature seems to be a measure of the attraction among particles
while its dependence on the parameter $b$ is more quantitative
then qualitative. On the other hand, as we have seen before if
$det(\eta_{ij})\rightarrow {0}$ then $R\rightarrow{\infty}$ i.e.
{\bf at the curve $\gamma_{\eta}$ of the signature change, scalar
curvature $R$ has the singularity inverse quadratic by the
distance to the curve}.
\par
\begin{remark}
It is interesting to note that, combining $(7.15)$ and Example 5
(see Sec. 7), we get
\par
\begin{equation}
(\frac{\partial \eta_{22}}{\partial
S})_{v}=\frac{\eta_{22}}{C_{v}}+\frac{2a}{C_{v}V^{3}},
\end{equation}.
\par
So, it seems reasonable to consider the quantity
\par
\begin{equation}
(\frac{\partial \eta_{22}}{\partial
S})_{v}-\frac{\eta_{22}}{C_{v}}=\delta(S,V)
\end{equation}
\par
as a measure of non-interaction of the system as long as $C_{v}$
is constant. In the case of the Ideal Gas, $\delta(S,V)=0$. For
the Van der Waals Gas, $\delta(S,V)=\frac{2a}{C_{v}V^{3}}$. notice
that the parameter $b$ is not present in $\delta(S,V)$. In the
limit where $a\rightarrow{0}$, with a fixed volume, Van der Waals
gas becomes Ideal even if $b\neq{0}$.
\end{remark}
\par
Let's now look at what happens when $R\rightarrow{\infty}$. As we
discussed earlier, when R goes to infinity, the system is
described by a \textit{degeneracy curves} $\gamma_{\eta}$ on which
phase transition seems to happen. Now, since the critical point is
obtained whenever both $(\frac{\partial p}{\partial v})_{T}=0$ and
$(\frac{\partial^{2}p}{\partial v^{2}})_{T}=0$ are satisfied, it
is evident that, the idea of phase transition is given by the
degeneracy of the Weinhold metric. What follows are the
derivations of the critical point for the Van der Waals gas
through the zero-determinant of the metric matrix $\eta_{U ij}$.
Indeed, considering the denominator of R being zero, we obtain:
\par
\begin{equation}
pV^{3}-aV+2ab=0
\end{equation}
\par
Then, we get the curve $\gamma_{\eta}$ in the p-V plane
\par
\begin{equation}
p(V)=(V-2b)\frac{a}{V^{3}}
\end{equation}
\par
Taking derivative with respect to V, we get
\par
\begin{equation}
\frac{dp}{dV}=(3b-V)\frac{2a}{V^4}
\end{equation}
\par
which is zero when
\par
\begin{equation}
V_{c}=V=3b
\end{equation}
\par
This is exactly the critical value of the volume for the van der
Waals gas. Now, substituting this value back into the equation
$(9.22)$, we get the critical value for pressure
\par
\begin{equation}
p_{c}=p=\frac{a}{27b^{2}}
\end{equation}
\par
Naturally, if we consider the denominator of the scalar curvature
with T and V in it and set it to zero we get the degeneracy curve
$\gamma_{\eta}$ in the T-V plane
\par
\begin{equation}
T(V)=\frac{2a(V-b)^{2}}{RV^{3}}
\end{equation}
\par
and since $V_{c}=3b$ then we get,
\par
\begin{equation}
T_{c}=\frac{8a}{27bR}
\end{equation}
\par
Consider, now, $p_{r}=\frac{p}{p_{c}}$, $V_{r}=\frac{V}{V_{c}}$
and $T_{r}=\frac{T}{T_{c}}$.
\par
Then
\par
\begin{equation}
p_{r}=\frac{3V_{r}-2}{V^{3}_{r}}
\end{equation}
\par
and
\par
\begin{equation}
T_{r}=\frac{(3V_{r}-1)^{2}}{4V^{3}_{r}}
\end{equation}
\par
Naturally, the two curves intersect at the critical point.
\par
Solving $(9.21)$ for $V_{r}$, we have the three roots of volume in
terms of pressure, namely
\par
\begin{equation}
V^{1}_{r}({p})=\frac{((3)^{\frac{1}{2}}sign(p)\cos(h(p))-\sin(h(p))}{p^{\frac{1}{2}}}
\end{equation}
\par
\begin{equation}
V^{2}_{r}({p})=-\frac{((3)^{\frac{1}{2}}sign(p)\cos(h(p))-\sin(h(p))}{p^{\frac{1}{2}}}
\end{equation}
\par
\begin{equation}
V^{3}_{r}({p})=\frac{2h(p)}{p^{\frac{1}{2}}}
\end{equation}
\par
where $h(p)=\frac{\arcsin(p^{\frac{1}{2}})}{3}.$
\par
Now, solving $(9.26)$ for $V_{r}$, we similarly get the roots of
volume in terms of temperature, namely
\par
\begin{equation}
V^{1}_{r}({T})=((3f(T))^{\frac{1}{2}}\cos(\frac{g(T)}{3})\frac{1}{4|T|}+(f(T)^{\frac{1}{2}}\sin(\frac{g(T)}{3})\frac{1}{4T}+\frac{3}{4T})
\end{equation}
\par
\begin{equation}
V^{2}_{r}({T})=-((3f(T))^{\frac{1}{2}}\cos(\frac{g(T)}{3})\frac{1}{4|T|}+(f(T)^{\frac{1}{2}}\sin(\frac{g(T)}{3})\frac{1}{4T}+\frac{3}{4T})
\end{equation}
\par
\begin{equation}
V^{3}_{r}({T})=\frac{3}{4T}-(f(T)^{\frac{1}{2}}\sin(\frac{g(T)}{3})\frac{1}{2T})
\end{equation}
\par
where $f(T)=9-8T $ and
\par
\begin{equation}
g(T)=\arctan(\frac{(8T^{2}-36T+27)(-f(T))^\frac{3}{2}}{8(f(T)^\frac{3}{2})(T^{3}(T-1))^\frac{1}{2}})\qquad\
\end{equation}
\par
%\begin{figure}
%\begin{center}
%\epsfig{figure=clausius.eps, height=2.0in, width=3.0in} \caption{
%Caption text. } \label{fig:1}
%\end{center}
%\end{figure}
%\par
%\begin{figure}
%\begin{center}
%\epsfig{figure=PT2.eps, height=2.0in, width=3.0in} \caption{
%Caption text. } \label{fig:PT2}
%\end{center}
%\end{figure}
%\par
%\begin{figure}
%\begin{center}
%\epsfig{figure=PT3.eps, height=2.0in, width=3.0in} \caption{
%Caption text. } \label{fig:PT3}
%\end{center}
%\end{figure}
\par
Therefore, substituting these last three solutions in $(9.28)$, we
get
\par
\begin{equation}
p^{i}_{r}({T})=\frac{3V^{i}_{r}({T})-2}{(V^{i}_{r}({T}))^{3}}
\end{equation}
\par
with $i=1,2,3$.
\par
In particular, for $i=3$, we get the interesting
\textit{coexistence} curve(Fig.1),
\par
\begin{figure}
\begin{center}
\epsfig{figure=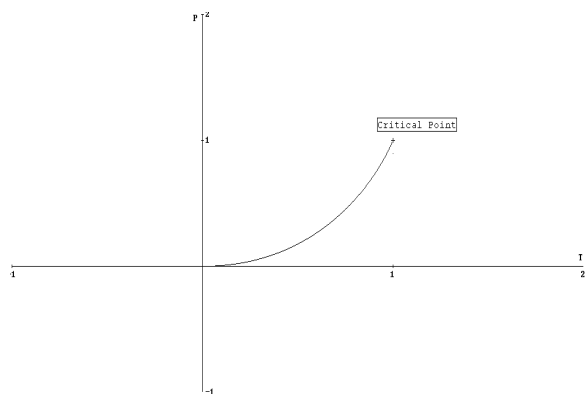, height=2.0in, width=3.0in}
\caption{Solution 1: Coexistence curve} \label{fig:clausius}
\end{center}
\end{figure}
\par
Considering $i=1,2$, instead, we get (Fig.2,3)\begin{figure}
\begin{center}
\epsfig{figure=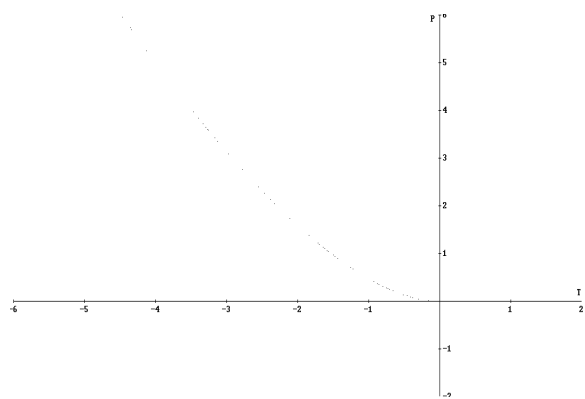, height=2.0in, width=3.0in}
\caption{Solution 2} \label{fig:PT2}
\end{center}
\end{figure}
\par
\begin{figure}
\begin{center}
\epsfig{figure=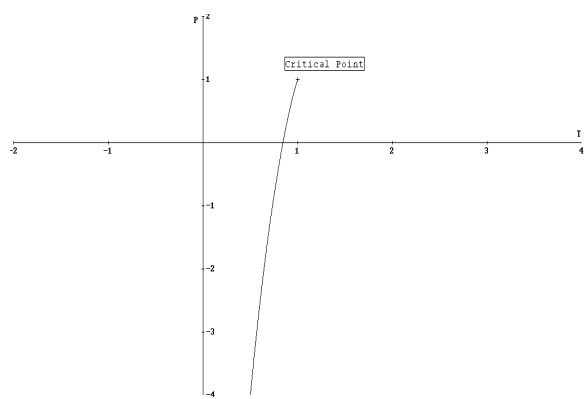, height=2.0in, width=3.0in}
\caption{Solution 3} \label{fig:PT3}
\end{center}
\end{figure}
\par
Now, differentiating $(9.28)$ and $(9.29)$, then we get
\par
\begin{equation}
\frac{dp_{r}}{dT_{r}}=\frac{8}{3V_{r}-1}
\end{equation}
\par
The important conclusion we gain by looking at the Van der Waals
gas is that the singular locus $Sing(\eta_{U})$ of Weinhold metric
carries important information on the critical behavior of the one
component system. In particular, along the curve
$Sing(\eta_{U})=\gamma_{\eta}$ phase transitions happens and
exterma of coordinate functions along this curve corresponds to
the critical points of the system.
\par
\mathstrut
\par
\section{Berthelot's gas}
\par
Consider Berthelot's gas (see \cite{KP}, Chapter 6) whose
conventional equation of state is
\par
\begin{equation}
(p+\frac{a}{TV^{2}})(V-b)=RT,
\end{equation}
where a and b are positive constants having the same meaning as
for the Van der Waals gas. For this gas, $C_{v}$ is not constant.
In particular,
\par
\begin{equation}
(C_{v})_{Bert}=(C_{v})_{ideal}+\frac{2a}{VT^{2}}
\end{equation}
\par
Moreover,
\par
\begin{equation}
k=\frac{TV^{2}(V-b)^{2}}{RT^{2}V^{3}-2a(V-b)^{2}}=\frac{TV^{2}(V-b)}{pTV^{3}-aV+2ab}
\end{equation}
\par
and
\par
\begin{equation}
\alpha=\frac{(V-b)(RT^{2}V^{2}+a(V-b))}{T(RT^{2}V^{3}-2a(V-b)^{2})}=\frac{(V-b)(pTV^{2}+2a)}{T(pTV^{3}-aV+2ab)}
\end{equation}
\par
Therefore the relation between $k$ and $\alpha$ is given by
\par
\begin{equation}
\frac{\alpha}{k}=\frac{R}{V-b}+\frac{a}{T^{2}V^{2}}
\end{equation}
\par
It is also interesting to look at the rates of change of the heat
capacity at constant volume. In particular, denoting
$(C_{v})_{Bert}$ by $C_{v}$ and considering $(10.2)$, we get
\par
\begin{equation}
(\frac{\partial C_{v}}{\partial S})_{v}=-\frac{4a}{C_{v}VT^{2}}
\end{equation}
\par
and
\par
\begin{equation}
(\frac{\partial C_{v}}{\partial
V})_{s}=\frac{2a}{C_{v}VT^{2}}(\frac{2a}{T^{2}V^{2}}+\frac{2R}{(V-b)}-\frac{C_{v}}{V})
\end{equation}
\par
The Weinhold metric $\eta_{ij}$ of the Berthelot's gas given by
\par
\begin{equation}
\eta_{U ij} =
\begin{pmatrix}
 \frac{T}{C_{v}} & -\frac{1}{C_{v}}(\frac{RT}{(V-b)}+\frac{a}{TV^{2}})\\
 -\frac{1}{C_{v}}(\frac{RT}{(V-b)}+\frac{a}{TV^{2}}) &
 \frac{RT^{2}V^{3}-2a(V-b)^{2}}{TV^{3}(V-b)^{2}}+\frac{T}{C_{v}}(\frac{R}{(V-b)}+\frac{a}{T^{2}V^{2}})^{2}
\end{pmatrix}
\end{equation}
\par
The determinant is then given by
\par
\begin{equation}
\det{(\eta_{U
ij})}=\frac{RT^{2}V^{3}-2a(V-b)^{2}}{C_{v}V^{3}(V-b)^{2}}
\end{equation}
\par
If we set the determinant of $\eta_{U ij}$ equal zero we get the
singular locus $Sing(\eta_{U})$ of the Bertellot gas
\par
\begin{equation}
RT^{2}V^{3}-2a(V-b)^{2}=2p^{2}V^{3}(V-b)^{2}-aR(V-2b)^{2}=0
\end{equation}
\par
 and solving the first equation for $T$ we get
\par
\begin{equation}
T(V)=\pm{\frac{(V-b)}{V}}(\frac{2a}{RV})^{\frac{1}{2}}
\end{equation}
\par
Taking derivative with respect to V, we get
\par
\begin{equation}
\frac{dT}{dV}=\pm{\frac{1}{2V}(\frac{2a}{RV})^{\frac{1}{2}}(2-\frac{3(V-b)}{V})}
\end{equation}
\par
which is zero when
\par
\begin{equation}
V_{c}=V=3b
\end{equation}
\par
considering $a\neq{0}$. This implies that
\par
\begin{equation}
T_{c}=\pm{(\frac{8a}{27Rb})^{\frac{1}{2}}}
\end{equation}
\par
and, therefore, since $p(V)=\frac{aV-2ab}{TV^{3}}$ whenever
$\det{(\eta_{U ij})}=0$, we have
\par
\begin{equation}
p_{c}=\pm{(\frac{aR}{216b^{3}})^{\frac{1}{2}}}
\end{equation}
\par
We could normalize both p and T as function of V. Then
\par
\begin{equation}
p^{2}_{r}=\frac{4(3V_{r}-2)^{2}}{V^{3}_{r}(3V_{r}-1)}
\end{equation}
\par
and
\par
\begin{equation}
T^{2}_{r}=\frac{3V_{r}-1}{4V^{3}_{r}}
\end{equation}
\par

Let's calculate the inverse matrix. We have
\par
\begin{equation}
\eta^{ij}_{U} =
\begin{pmatrix}
 \frac{C_{v}}{T}+\frac{RV(RT^{2}V^{2}+a(V-b))}{T(RT^{2}V^{3}-2a(V-b)^{2})}+\frac{a(V-b)(RT^{2}V^{2}+a(V-b))}{VT^{3}(RT^{2}V^{3}-2a(V-b)^{2})} & \frac{V(V-b)(RT^{2}V^{2}+a(V-b))}{T(RT^{2}V^{3}-2a(V-b)^{2})}\\
 \frac{V(V-b)(RT^{2}V^{2}+a(V-b))}{T(RT^{2}V^{3}-2a(V-b)^{2})} &
 \frac{TV^{3}(V-b)^{2}}{RT^{2}V^{3}-2a(V-b)^{2}}
\end{pmatrix}
\end{equation}
\par
It is convenient, now, to calculate the third derivatives without
grouping, in particular,
\par
\begin{equation}
\eta_{{11,}_{1}}=\frac{T}{C^{2}_{v}}+\frac{4a}{VTC^{3}_{v}}
\end{equation}
\par
\begin{equation}
\eta_{{12,}_{1}}=\eta_{{21,}_{1}}=\eta_{{11,}_{2}}=-\frac{R}{(V-b)}\eta_{{11,}_{1}}+\frac{a}{C^{2}_{v}TV^{2}}(1+(\frac{\partial{C_{v}}}{\partial{S}})_{v})
\end{equation}
\par
\[
\eta_{{12,}_{2}}=\eta_{{21,}_{2}}=\eta_{{22,}_{1}}=\frac{R^{2}}{(V-b)^{2}}\eta_{{11,}_{1}}-\frac{2aR}{C^{2}_{v}TV^{2}(V-b)}(1+(\frac{\partial{C_{v}}}{\partial{S}})_{v})+\frac{RT}{C_{v}(V-b)^{2}}
\]
\par
\begin{equation}
+\frac{a}{C_{v}TV^{3}}[\frac{4a^{2}}{C^{2}_{v}T^{4}V^{2}}-\frac{3a}{C_{v}VT^{2}}+2]
\end{equation}
\par
\[
\eta_{{22,}_{2}}=-\frac{R^{3}}{(V-b)^{3}}\eta_{{11,}_{1}}+\frac{\alpha}{k}[\frac{12a^{2}R}{T^{3}V^{3}C^{3}_{v}(V-b)}+\frac{7aR}{TV^{2}C^{2}_{v}(V-b)}+\frac{2aR}{TC^{2}_{v}}-\frac{6a}{C_{v}TV^{3}}-\frac{3RT}{C_{v}(V-b)^{2}}]
\]
\par
\begin{equation}
-\frac{2RT}{(V-b)^{2}}[\frac{3aR}{T^{2}V^{2}C^{2}_{v}}-\frac{1}{(V-b)}]-\frac{a}{TV^{4}}[\frac{4a^{3}}{C^{3}_{v}T^{6}V^{3}}-\frac{5a^{2}}{C^{2}_{v}T^{4}V^{2}}+6]
\end{equation}
\par
Then, from $(6.27)$, we get the scalar curvature to be
\par
\begin{equation}
R(\eta_{U})=2a\frac{\left(T^{4}V^{4}RC_{v}P(C_{v},V)+T^{2}V^{3}RaQ(C_{v},V)+a^{2}W(C_{v},V)\right)}{C^{3}_{v}T^{3}V(RT^{2}V^{3}-2a(V-b)^{2})^{2}},
\end{equation}
\par
where
\par
\[
P(C_{v},V)=(2C_{v}-R)V^{2}-3C_{v}bV+C_{v}b^{2}
\]
\par
\[
Q(C_{v},V)=-RV^{5}+3RbV^{4}-3Rb^{2}V^{3}+(Rb^{3}+C_{v}+R)V^{2}-b(b-2V)(R+C_{v})
\]
\par
and
\par
\[
W(C_{v},V)=-RV^{7}+4RbV^{6}-6Rb^{2}V^{5}+(2C_{v}+R+4Rb^{3})V^{4}-(8C_{v}+3R+Rb^{3})bV^{3}+(12C_{v}+3R)b^{2}V^{2}-
\]
\par
\[
-(8C_{v}+R)b^{3}V+2C_{v}b^{4}
\]
Notice that, as for VdW gas, as $a$ goes to zero then scalar
curvature $R$ goes to zero as well, while it is not true in the
case of $b$ being zero.

\section{Geodesic equations}
\par
We found the Christoffel coefficients for a general 2-dimensional
thermodynamical phase space to be
\par
\begin{equation}
\Gamma^{k}_{ij}=\frac{1}{2}\sum_{m}\eta_{ij,_{m}}\eta^{km}
\end{equation}
\par
Now, the geodesic equations in a general form are given by
\par
\begin{equation}
\frac{d^{2}x^{k}}{dt^{2}}+\Gamma^{k}_{ij}\frac{dx^{i}}{dt}\frac{dx^{j}}{dt}=0
\end{equation}
\par
For our case in which $E=U(S,V)$, the geodesic equations are of
the form
\par
\begin{equation}
\frac{d^{2}S}{dt^{2}}+\Gamma^{1}_{11}(\frac{dS}{dt})^{2}+2\Gamma^{1}_{12}\frac{dS}{dt}\frac{dV}{dt}+\Gamma^{1}_{22}(\frac{dV}{dt})^{2}=0
\end{equation}
\par
\begin{equation}
\frac{d^{2}V}{dt^{2}}+\Gamma^{2}_{11}(\frac{dS}{dt})^{2}+2\Gamma^{2}_{12}\frac{dS}{dt}\frac{dV}{dt}+\Gamma^{2}_{22}(\frac{dV}{dt})^{2}=0
\end{equation}
\par
Let's look at the Christoffel coefficients in more details.
\par
\begin{equation}
\Gamma^{1}_{11}=\frac{1}{2}(\eta_{11,_{1}}\eta^{11}+\eta_{11,_{2}}\eta^{12})\qquad
\Gamma^{2}_{11}=\frac{1}{2}(\eta_{11,_{1}}\eta^{21}+\eta_{11,_{2}}\eta^{22})
\end{equation}
\par
\begin{equation}
\Gamma^{1}_{12}=\Gamma^{1}_{21}=\frac{1}{2}(\eta_{12,_{1}}\eta^{11}+\eta_{12,_{2}}\eta^{12})\qquad
\Gamma^{2}_{12}=\Gamma^{2}_{21}=\frac{1}{2}(\eta_{12,_{1}}\eta^{21}+\eta_{12,_{2}}\eta^{22})
\end{equation}
\par
\begin{equation}
\Gamma^{1}_{22}=\frac{1}{2}(\eta_{22,_{1}}\eta^{11}+\eta_{22,_{2}}\eta^{12})\qquad
\Gamma^{2}_{22}=\frac{1}{2}(\eta_{22,_{1}}\eta^{21}+\eta_{22,_{2}}\eta^{22})
\end{equation}
\par
Considering the following relations:
\par
\begin{equation}
F=k(\frac{\partial}{\partial{V}}\ln{\frac{k}{\alpha}})_{S}
\end{equation}
\par
\begin{equation}
J=1-(\frac{\partial{C_{v}}}{\partial{S}})_{V}
\end{equation}
\par
\begin{equation}
D=\frac{\alpha}{k}+(\frac{\partial{C_{v}}}{\partial{V}})_{s}
\end{equation}
\par
and
\par
\begin{equation}
B=\frac{\alpha}{V}+(\frac{\partial{\alpha}}{\partial{V}})_{s}
\end{equation}
\par
then, we can calculate the coefficients explicitly,
\par
\begin{equation}
\Gamma^{1}_{11}=\frac{1}{2}[\frac{C_{p}}{C^{2}_{v}}J-\frac{TV\alpha}{C^{2}_{v}}D]
\end{equation}
\par
\begin{equation}
\Gamma^{1}_{12}=\Gamma^{1}_{21}=-\frac{1}{2}[\frac{1}{C_{v}}D+\frac{TV\alpha^{2}}{k^{2}C_{v}}F]
\end{equation}
\par
\begin{equation}
\Gamma^{1}_{22}=\frac{1}{2}[\frac{\alpha}{kC_{v}}D-\frac{TV\alpha^{3}}{k^{3}C_{v}}F-\frac{1}{k}B]
\end{equation}
\par
\begin{equation}
\Gamma^{2}_{11}=\frac{1}{2}[\frac{TV\alpha}{C^{2}_{v}}J-\frac{TVk}{C^{2}_{v}}D]
\end{equation}
\par
\begin{equation}
\Gamma^{2}_{12}=\Gamma^{2}_{21}=\frac{1}{2}[\frac{TV\alpha}{kC_{v}}F]
\end{equation}
\par
\begin{equation}
\Gamma^{2}_{22}=-\frac{1}{2}[\frac{C_{p}}{kC_{v}}F+\frac{1}{\alpha}B]
\end{equation}
\par
Therefore we have the following proposition:
\begin{proposition}
The geodesic equations for a general two dimensional
thermodynamical system are given by
\par
\begin{equation}
\frac{d^{2}S}{dt^{2}}+\frac{1}{2}[\frac{C_{p}}{C^{2}_{v}}J-\frac{TV\alpha}{C^{2}_{v}}D](\frac{dS}{dt})^{2}-[\frac{1}{C_{v}}D+\frac{TV\alpha^{2}}{k^{2}C_{v}}F]\frac{dS}{dt}\frac{dV}{dt}+\frac{1}{2}[\frac{\alpha}{kC_{v}}D-\frac{TV\alpha^{3}}{k^{3}C_{v}}F-\frac{1}{k}B](\frac{dV}{dt})^{2}=0
\end{equation}
\par
\begin{equation}
\frac{d^{2}V}{dt^{2}}+\frac{1}{2}[\frac{TV\alpha}{C^{2}_{v}}J-\frac{TVk}{C^{2}_{v}}D](\frac{dS}{dt})^{2}+[\frac{TV\alpha}{kC_{v}}F]\frac{dS}{dt}\frac{dV}{dt}-\frac{1}{2}[\frac{C_{p}}{kC_{v}}F+\frac{1}{\alpha}B](\frac{dV}{dt})^{2}=0
\end{equation}
\end{proposition}
\par
In the case in which $C_{v}$ is constant, then $B$ and $F$ are the
same while the other relations become
\par
\begin{equation}
G=0
\end{equation}
\par
\begin{equation}
J=1
\end{equation}
\par
\begin{equation}
D=\frac{\alpha}{k}
\end{equation}
\par
So, now, our Christoffel's coefficients become
\par
\begin{equation}
\Gamma^{1}_{11}=\frac{1}{2C_{v}}
\end{equation}
\par
\begin{equation}
\Gamma^{1}_{12}=\Gamma^{1}_{21}=-\frac{1}{2}[\frac{\alpha}{kC_{v}}+\frac{TV\alpha^{2}}{k^{2}C_{v}}F]
\end{equation}
\par
\begin{equation}
\Gamma^{1}_{22}=\frac{1}{2}[\frac{\alpha^{2}}{k^{2}C_{v}}-\frac{TV\alpha^{3}}{k^{3}C_{v}}F-\frac{1}{k}B]
\end{equation}
\par
\begin{equation}
\Gamma^{2}_{11}=0
\end{equation}
\par
\begin{equation}
\Gamma^{2}_{12}=\Gamma^{2}_{21}=\frac{1}{2}[\frac{TV\alpha}{kC_{v}}F]
\end{equation}
\par
\begin{equation}
\Gamma^{2}_{22}=-\frac{1}{2}[\frac{C_{p}}{kC_{v}}F+\frac{1}{\alpha}B]
\end{equation}
\par
So, finally, our geodesic equation with $C_{v}$ constant are given
by
\par
\begin{lemma}
\par
\begin{equation}
\frac{d^{2}S}{dt^{2}}+\frac{1}{2C_{v}}(\frac{dS}{dt})^{2}-[\frac{\alpha}{kC_{v}}+\frac{TV\alpha^{2}}{k^{2}C_{v}}F]\frac{dS}{dt}\frac{dV}{dt}+\frac{1}{2}[\frac{\alpha^{2}}{k^{2}C_{v}}-\frac{TV\alpha^{3}}{k^{3}C_{v}}F-\frac{1}{k}B](\frac{dV}{dt})^{2}=0
\end{equation}
\par
\begin{equation}
\frac{d^{2}V}{dt^{2}}+[\frac{TV\alpha}{kC_{v}}F]\frac{dS}{dt}\frac{dV}{dt}-\frac{1}{2}[\frac{C_{p}}{kC_{v}}F+\frac{1}{\alpha}B](\frac{dV}{dt})^{2}=0
\end{equation}
\end{lemma}
\par
\section{Conclusion}
\par
In this work we have studied the scalar (Gauss) curvature of
Weinhold metric for a thermodynamical systems with two
thermodynamical degrees of freedom. We get criteria for the
positivity, nullity and negativity of scalar curvature in terms of
{\it Hessian surface} of the thermodynamical potential, found
scalar curvature for a general thermodynamical systems with two
thermodynamical degrees of freedom.  We have studied relation of
the signature change of Weinhold metric and the scalar curvature
to the curves of phase transition of these systems.  As examples
we have considered the systems with the heat capacity $C_{v}$
constant, in particular the Ideal and Van der Waals gases, and the
Berthelot gas.  Results obtained here suggest a kind of duality
relation between the constitutive surface of a 2D thermodynamical
system in the Gibbs space (Space with coordinates $(U,S,V)$ in the
case of internal energy) and its Hessian surface.  Relations
between the convexity properties of both surfaces, curvature and
signature of thermodynamical metric, extremal properties of
corresponding thermodynamical potential and the phase transitions
in the thermodynamical system present interesting and, in our
opinion, highly promising direction of the future work.
\par
\mathstrut

\end{document}